\documentclass[10pt]{article}
\usepackage{geometry}
\usepackage{xcolor}
\definecolor{graycolor}{gray}{0.9} 
\usepackage{microtype}
\usepackage{setspace} 
\usepackage[utf8]{inputenc}
\usepackage[english]{babel}
\usepackage{times}
\usepackage{array}
\usepackage{soul}
\usepackage{amsfonts,amsmath,amssymb}
\usepackage{latexsym,color,cite} 
\usepackage{cite}
\usepackage[numbers,sort&compress]{natbib}
\usepackage{natbib}

\usepackage{titlesec} 
\titleformat {\section} [block] {\raggedright \fontsize{10}{10}\selectfont\bfseries} {\thesection. \space} {0pt} {}
\titlespacing {\section} {0pt} {12pt} {6pt}
\titleformat {\subsection} [block] {\raggedright \fontsize{10}{10}\selectfont\itshape} {\thesubsection .\space} {0pt} {}
\titlespacing {\subsection} {0pt} {12pt} {6pt}
\titleformat {\subsubsection} [block] {\raggedright \fontsize{10}{10}\selectfont} {\thesubsubsection .\space} {0pt} {}
\titlespacing {\subsubsection} {0pt} {12pt} {6pt}
\titleformat {\paragraph} [block] {\raggedright \fontsize{10}{10}\selectfont} {} {0pt} {}
\titlespacing {\paragraph} {0pt} {12pt} {6pt}

\usepackage{array} \newcommand{\PreserveBackslash}[1]{\let\temp=\\#1\let\\=\temp}
\newcolumntype{C}[1]{>{\PreserveBackslash\centering}m{#1}}
\newcolumntype{R}[1]{>{\PreserveBackslash\raggedleft}m{#1}}
\newcolumntype{L}[1]{>{\PreserveBackslash\raggedright}m{#1}}
\usepackage{lineno}
\usepackage{tabularx}
\usepackage{colortbl}
\usepackage{graphicx}
\usepackage{float}
\usepackage[export]{adjustbox}
\usepackage{caption}
\usepackage{fancyhdr} 
\def\eqn#1{\eq~\eqref{#1}}
\def\rf{\eqref}
\def\mn{_{\mu\nu}}
\def\MN{^{\mu\nu}}
\def\mN{_\mu^\nu}

\def\D{\partial}

\def\R{{\mathbb R}}

\def\cF{{\mathcal F}}

\def\og{{\overline g}}

\def\oR{{\overline R}}
\def\dxi{\delta\xi}
\def\Veff{V_{\rm eff}}
\def\oalpha{\overline{\alpha}}
\def\obeta{\overline{\beta}}
\def\ogamma{\overline{\gamma}}
\def\chg{\ \leftrightarrow\ }

\def\sph{spherically symmetric}
\def\ssph{static, spherically symmetric}
\def\bh{black hole}
\def\bhs{black holes}
\def\wh{wormhole}

\def\asflat{asymptotically flat}
\def\emag{electromagnetic}

\def\mult{multidimensional}
\def\pb{perturbation}
\def\pbs{perturbations}
\def\Scw{Schwarz\-schild}
\def\Schr{Schr\"odinger}
\def\RN{Reiss\-ner-Nord\-str\"om}

\def\ApJ#1 {\emph{Astroph. J.}~{#1}~}
\def\CQG#1 {\emph{Class. Quantum Grav.}~{#1}~}
\def\DAN#1 {\emph{Dokl. AN SSSR}~{#1}~}
\def\GC#1 {\emph{Grav. Cosmol.}~{#1}~}
\def\GRG#1 {\emph{Gen. Rel. Grav.}~{#1}~}
\def\IJMPD#1 {\emph{Int. J. Mod. Phys. D}~{#1}~}
\def\JETF#1 {\emph{Zh. Eksp. Teor. Fiz.}~{#1}~}
\def\JETP#1 {\emph{Sov. Phys. JETP}~{#1}~}
\def\JHEP#1 {\emph{JHEP}~{#1}~}
\def\JMP#1 {\emph{J. Math. Phys.}~{#1}~}
\def\NPB#1 {\emph{Nucl. Phys. B}~{#1}~}
\def\NP#1 {\emph{Nucl. Phys.}~{#1}~}
\def\PLA#1 {\emph{Phys. Lett. A}~{#1}~}
\def\PLB#1 {\emph{Phys. Lett. B}~{#1}~}
\def\PRD#1 {\emph{Phys. Rev. D}~{#1}~}
\def\PRL#1 {\emph{Phys. Rev. Lett.}~{#1}~}

\def\lal{&&\nqq {}}
\def\eq{Equation\,}

\def\beq{\begin{equation}}
\def\eeq{\end{equation}}
\def\bear{\begin{eqnarray}}
\def\bearr{\begin{eqnarray} \lal}
\def\ear{\end{eqnarray}}
\def\earn{\nonumber \end{eqnarray}}

\def\nnn{\nonumber\\ \lal }

\def\yy{\\[5pt] {}}
\def\yyy{\\[5pt] \lal }

\def\e{{\,\rm e}}

\def\im{\mathop{\rm Im}\nolimits}
\def\arg{\mathop{\rm arg}\nolimits}

\def\sign{\mathop{\rm sign}\nolimits}
\def\diag{\mathop{\rm diag}\nolimits}

\def\const{{\rm const}}

\def\ep{\epsilon}

\def\then{\ \Rightarrow\ }
\newcommand{\toas}{\mathop {\ \longrightarrow\ }\limits }

\newcommand{\vars}[1]{\left\{\begin{array}{ll}#1\end{array}\right.}

\def\nqq{\hspace*{-2em}}

\def\qq{\qquad}
\def\cm{\hspace*{1cm}}
\def\inch{\hspace*{1in}}

\def\dst{\displaystyle}
\def\tst{\textstyle}
\def\fracd#1#2{{\dst\frac{#1}{#2}}}
\def\fract#1#2{{\tst\frac{#1}{#2}}}
\def\Half{{\fracd{1}{2}}}
\def\half{{\fract{1}{2}}}

\usepackage{lastpage}
\usepackage{layout}
\usepackage{setspace} 
\usepackage{enumitem}
\usepackage{booktabs}
\usepackage{arydshln}
\usepackage{multirow}
\usepackage{color}
\usepackage{hyperref} 
\hypersetup{
	colorlinks=true,
	linkcolor=blue,
	filecolor=blue,
	urlcolor=black,
	citecolor=cyan,
}

\setcitestyle{open={[},close={]},citesep={,\!},numbers}

\fancypagestyle{firstpage}{
    \setlength{\headsep}{2.2cm}
    
    \setlength{\footskip}{1.5cm}
    \fancyhf{}
    \lhead{\begin{table}[H]
        \centering
        \begin{tabular}{L{2.5cm}C{10cm}C{3.1cm}R{2cm}}
            \includegraphics[scale=0.035]{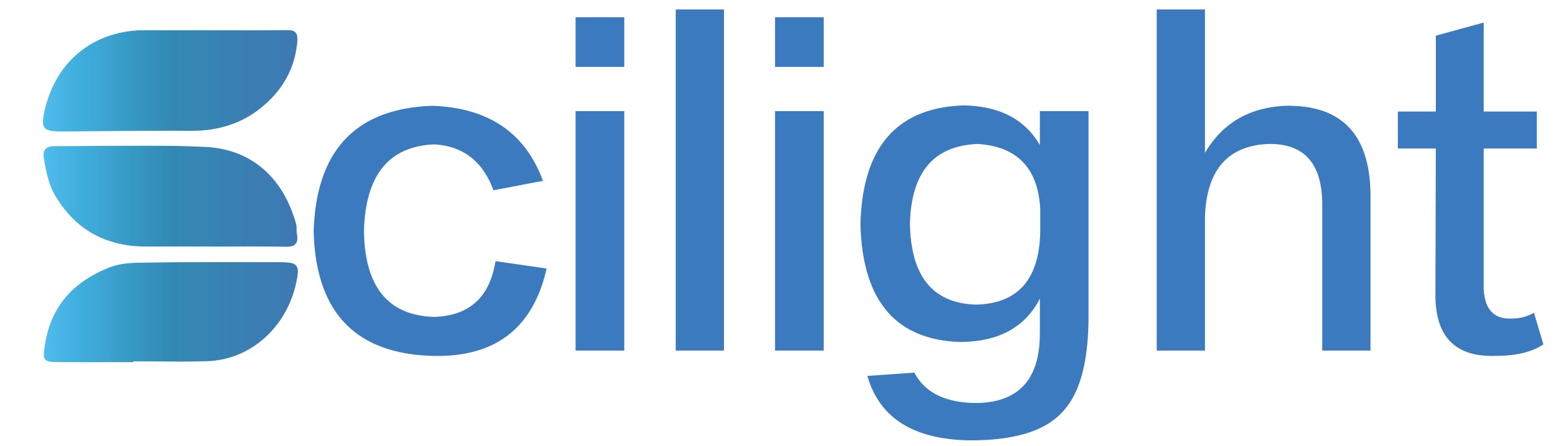} \vspace{-6pt}& \cellcolor{graycolor}\begin{tabular}[c]{@{}c@{}}\textit{International Journal of Gravitation and Theoretical Physics}\\ \href{https://www.sciltp.com/journals/ijgtp}{https://www.sciltp.com/journals/ijgtp}\end{tabular} & \includegraphics[scale=0.0150]{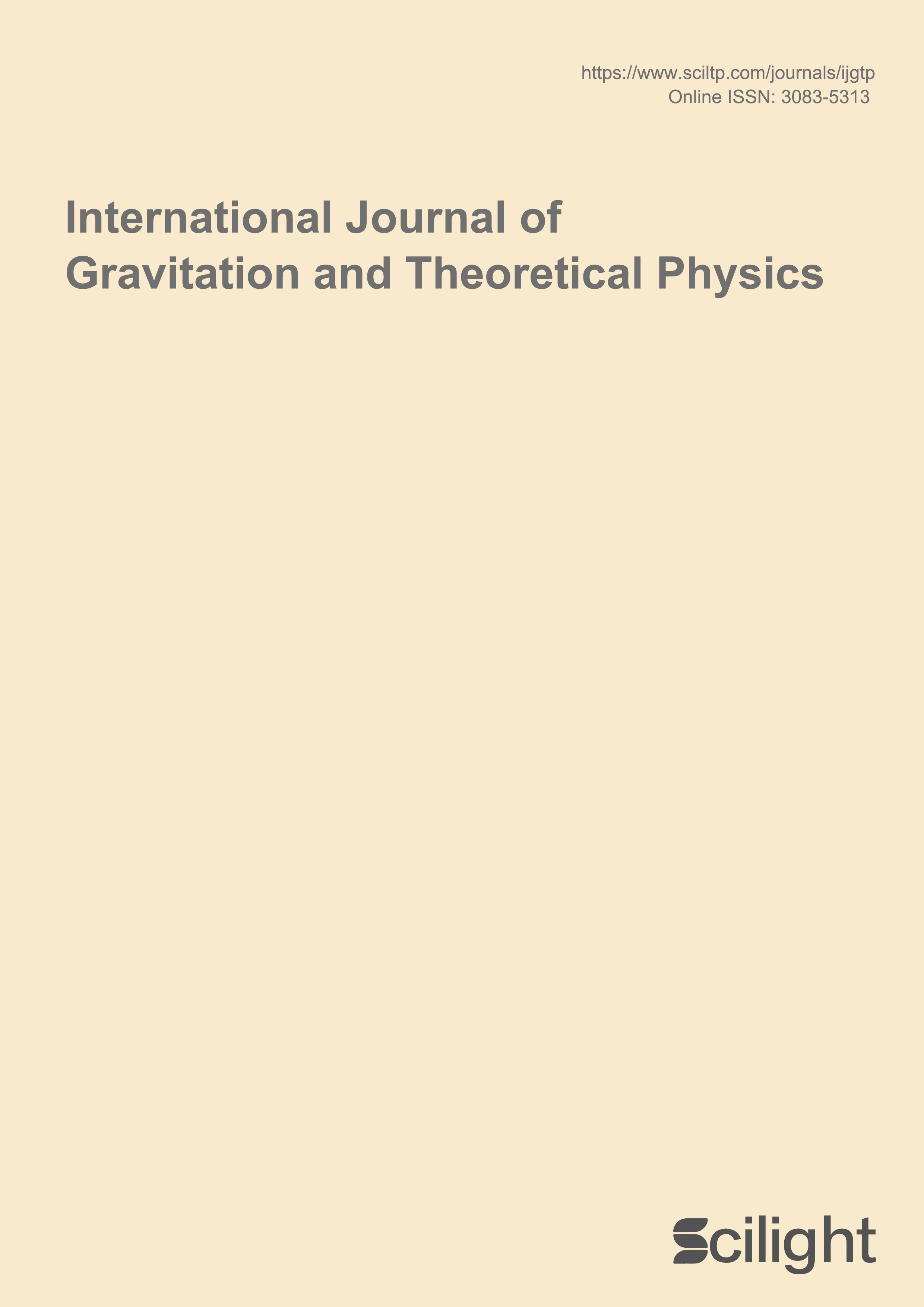} \vspace{-3pt}\\
        \end{tabular}
        \vspace{-22pt}
    \end{table}}
   
    \fancyfoot[C]{
        \vspace{-1.55cm}
        \begin{table}[H]
            \begin{minipage}[c]{0.15\columnwidth}
                \includegraphics[scale=0.5]{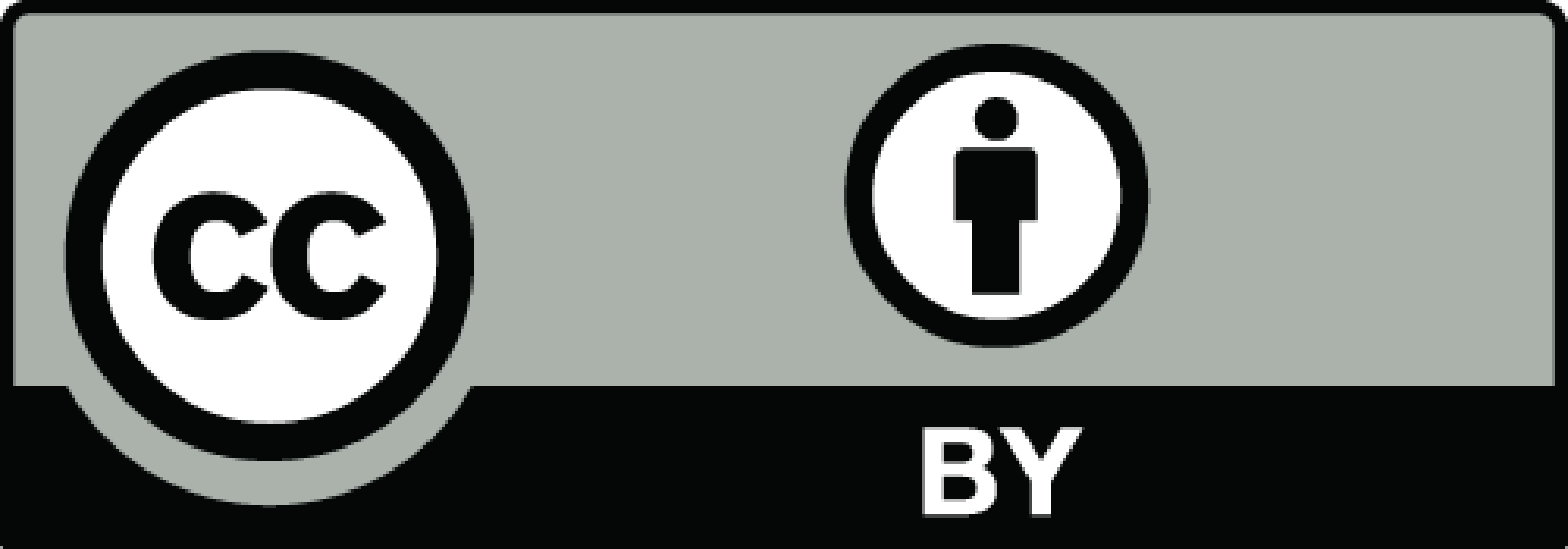} \vspace{1.1pt}
            \end{minipage}
            \hfill
            \begin{minipage}[c]{0.85\columnwidth}
                \scriptsize \textbf{Copyright:} © 2026 by the authors. This is an open access article under the terms and conditions of the Creative Commons Attribution (\mbox{CC BY}) license (\href{https://creativecommons.org/licenses/by/4.0/}{https://creativecommons.org/licenses/by/4.0/}). \\ \textbf{Publisher’s Note:} Scilight stays neutral with regard to jurisdictional claims in published maps and institutional affiliations.
            \end{minipage}
    \end{table}}
    \vspace{-0.55cm}
}

\begin{document}
\newgeometry{left=2.5cm, right=2.5cm, top=1.8cm, bottom=4cm}
	\thispagestyle{firstpage}
	\nolinenumbers
	{\noindent \textit{Article
    }}
	\vspace{4pt} \\
	{\fontsize{18pt}{10pt}\textbf{Stability Ranges of Magnetic Black Holes and Mirror\\ (\mbox{Topological}) Stars in 5D Gravity}}
	\vspace{16pt} \\
	{\large Kirill A. Bronnikov \textsuperscript{1,2,3}, Sergei V. Bolokhov \textsuperscript{2,*} and Milena V. Skvortsova \textsuperscript{2} }
	\vspace{6pt}
	 \begin{spacing}{0.9}
		{\noindent \small
			\textsuperscript{1}	Center of Gravitation and Fundamental Metrology, Rostest, 
		Ozyornaya ul. 46, Moscow 119361, Russia \\
			\textsuperscript{2}	Institute of Gravitation and Cosmology, RUDN University, 
		ul. Miklukho-Maklaya 6, Moscow 117198, Russia \\
			\textsuperscript{3}	National Research Nuclear University ``MEPhI'', 
		Kashirskoe sh. 31, Moscow 115409, Russia \\
        		    {*}  \parbox[t]{0.98\linewidth}{Correspondence: bolokhov-sv@rudn.ru} 	\vspace{6pt}\\
		\footnotesize	\textbf{How To Cite}: Bronnikov, K.A.; Bolokhov, S.V.; Skvortsova, M.V. Stability Ranges of Magnetic Black Holes and Mirror ({Topological}) Stars in 5D Gravity. \emph{International Journal of Gravitation and Theoretical Physics} \textbf{2026}, \emph{2}(1), 2. \href{https://doi.org/10.53941/ijgtp.2026.100002}{https://doi.org/10.53941/ijgtp.2026.100002}}\\
	\end{spacing}

\begin{table}[H]
\noindent\rule[0.15\baselineskip]{\textwidth}{0.5pt} 
\begin{tabular}{lp{12cm}}  
 \small 
  \begin{tabular}[t]{@{}l@{}} 
  \end{tabular} &
  \textbf{Abstract:} We discuss static, spherically symmetric solutions to the 5D Einstein-Maxwell equations 
    (belonging to wide classes of multidimensional solutions known at least from the 1990s)
    and select among them those which must observationally look like local objects whose
    surface reflects back particles or signals getting there, the so-called mirror stars
    (also called ``topological stars'' by some authors). Their significant parameters are the
    Schwarzschild mass $m$ and the magnetic charge $q$, such that $q^2 > 3m^2$, 
    {while the radius of their mirror surface is $r_b = 2q^2/(3m) > 2m$.}
    We also discuss their black hole counterparts for which $q^2 \leq 3m^2$. For both these
    objects, we study spherically symmetric time-dependent perturbations and determine the 
    stability regions in their parameter spaces. {Thus, mirror stars turn out to be stable 
    only at $r_b < r_b^{\rm crit} \approx 4.004\,m$, while the black holes prove to be stable in 
    the whole range of their parameters. We calculate the fundamental frequencies
    and decay rates of black hole perturbations using the WKB and time domain methods.}
    Our stability results disagree with some of those previously announced in the literature.  \\
\\
  & 
\end{tabular}
\noindent\rule[0.15\baselineskip]{\textwidth}{0.5pt} 
\end{table}

\section{Introduction }
	Multidimensional gravity contains a great variety of possible space-time configurations, both 
   evidently imaginary and potentially realistic ones, both in the cosmological context and in  
   modeling compact objects with strong gravitational fields. Besides a rich collection of 
   \bh\ models (see, e.g., \cite{bh1, bh2, bh3, bh4, bh5} for reviews), there are \mult\ models of
   wormholes \cite{wh1, wh2, wh3, wh4, wh5}, boson stars \cite{bos1, bos2, bos3}, gravastars 
   \cite{gras1, gras2}, etc. that possess many features of interest but are still extensions of 
   certain known 4D objects.
   
   Unlike that, in the present paper we would like to discuss such hypothetic objects whose 
   very existence becomes possible due to extra dimensions. They are still related to
   multidimensional \bh\ solutions in the following way: consider such a \bh\ metric and 
   mutually substitute one of the extra coordinates and the original time coordinate 
   \cite{kb95a, kb95b}. Such a replacement surely leads to a new solution since for the 
   equations it does not matter which of the coordinates is interpreted as time and which 
   is regarded ``extra.'' It can be inferred that such solutions are almost as numerous as 
   are \bh\ ones: for their existence, it is only required that among the extra dimensions 
   there is a suitable 1D subspace. As argued in  \cite{kb95a, kb95b}, if this extra 1D subspace 
   is compact and sufficiently small to be invisible by modern instruments, then the surface 
   that had been an event horizon in the original \bh\ solution, now becomes a perfectly
   reflecting surface. We therefore proposed to call such objects {\sl mirror stars} \cite{we25}.
   
    Various reflection phenomena in astrophysics are rather actively discussed in the
   literature. Different kinds of echoes have been predicted in \bh\ and \wh\ space-times
   \textls[-15]{\cite{ech1,ech2,ech3,ech4}. In particular, the authors of \cite{25-lim} have obtained explicit
    observational constraints on possible reflective compact objects whose surface radius $r_s$
 is close to the would-be event horizon radius $r_h$, $r_s = r_h (1 + \ep)$. According to
   \cite{25-lim}, such objects with the \mbox{``compactness  parameter''}}
   
   \restoregeometry

\noindent  $1 + \ep < 1 + 10^{-3}$ are almost excluded.
   
   In what follows we will discuss some simple examples of mirror star solutions to the 5D
   Einstein-Maxwell equations, which are special cases of \ssph\ solutions obtained in 
   \cite{kb95a, kb95b}. Among them, the solution with a magnetic charge seems to be
   of largest interest, even though the smallness of extra dimensions, required for making 
   them invisible by our instruments, leads to a severe restriction on mirror star masses, at 
   least according to the solutions so far obtained. For comparison we also discuss the 
   corresponding 5D \bh\ (BH) solutions, which also possess some features of interest. 
    
   We study their stability under \sph\ \pbs\ and determine the stability ranges of their 
   parameters. According to our results, for stable mirror stars, the above-mentioned 
   ``compactness parameter'' belongs to 
   the range {$1 + \ep \in (1, 2.002)$} (with a numerically obtained upper bound). 
{Meanwhile, the black-hole solutions turn out to be stable in the whole range of their
   parameters, and we obtain estimates for the decay characteristics of their perturbations.}
      
   It should be noted that the objects that we call ``magnetic mirror stars'' have been recently
   discussed under the name of ``topological stars'' \cite{tops1, tops2, tops3}, and in 
   \cite{tops3} they were found to be stable under nonspherical \pbs. However, the stability   
   range under \sph\ \pbs\ stated there does not coincide with ours, making necessary a
   further study. The methodologies of obtaining the solutions in question and their stability
   studies in  \cite{tops1, tops2, tops3} are different from ours, and therefore we believe that
   this comparison is of particular interest.
        
   This paper is organized as follows. In Section \ref{sec2} we consider the simplest example
   of a mirror-star solution being a 5D analogue of the Schwarzschild metric. 
   Section \ref{sec3} considers 5D Einstein-Maxwell fields, and it is concluded that among them 
   the magnetic solutions of both \bh\ and mirror star types are of utmost interest. 
   Their stability properties under \sph\ \pbs\ are studied in Section \ref{sec-stab}, and Section \ref{sec5} is a~conclusion.
   
\section{Mirror Stars: A Simple Example}\label{sec2}

  As mentioned above, in \mult\ space-times, in addition to BH solutions, there are families of 
  so-called T-hole ones \cite{kb95a, kb95b}, formally obtained from BH ones by substituting
  $t\chg v$, where $v$ is one of the extra coordinates. However, they possess certain features
  connected with compactness of the extra dimensions, which are most clearly understood
  considering the simplest example on the basis of \Scw's solution.
  
  We will deal with 5D general relativity (GR) and \ssph\ metrics of the form
\beq           \label{ds5}
		ds_5^2 = \e^{2\gamma} dt^2 - \e^{2\alpha} du^2 - \e^{2\beta} d\Omega^2 
				+\eta_v \e^{2\xi} dv^2,
\eeq    
  where $u$ is the radial coordinate (admitting an arbitrary parametrization), $v$ is 
  the fifth coordinate, and $\eta_v = \pm 1$ depending on whether $v$ is timelike ($+1$) 
  or spacelike ($-1$). 

  One of the vacuum solutions of this theory is the extended Schwarzschild solution, 
  representing the \Scw\ metric in its usual form with the added term $\eta_v dv^2$. 
  The field equations ``do not know'' which of the coordinates is observable time and 
  which is the extra one, hence there is another solution (to be called {\it T-Schwarzschild})
  with the~metric
\beq	  \label{TScw}
	 ds_5^2 = dt^2 - \bigg(1- \frac{2m}{r}\bigg)^{-1} dr^2 - r^2 d\Omega^2
	 			+ \bigg(1- \frac{2m}{r}\bigg)\eta_v dv^2;
\eeq
  when crossing the value $r=2m$, both signs of $g_{rr}$ and $g_{vv}$ simultaneously
  change. If $\eta_v =1$, i.e., if the $v$ direction is timelike at large $r$, the overall 
  signature of the 5D metric is preserved, but in the opposite case, $\eta_v =-1$, it changes 
  by four since two spacelike directions become timelike. Meanwhile, as can be directly 
  verified, such a horizon is not a curvature singularity, both in the 5D metric and 
  in its 4D section (The 4D part of \rf{TScw}, if we forget about the extra subspace,
  describes the so-called \Scw\ traversable wormhole, in which $r=2m$ is the~throat).

\begin{figure} [H]   
\centering     
\includegraphics[width=7cm]{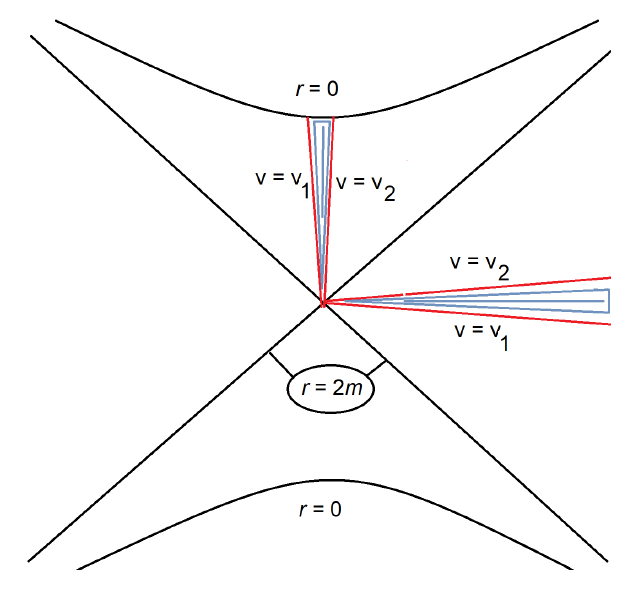}
\caption{\small
	The Kruskal diagram for the metric \rf{TScw} with a temporal extra dimension.}\label{fig1}
\end{figure}

  If $\eta_v=1$, the surface $r = 2m$ looks like a conventional \Scw\ horizon in the
  $(R,v)$ subspace, admitting an analytic extension from $r> 2m$ to $r < 2m$ with the
  corresponding Kruskal picture. However, if some points on the $v$ axis are identified, as 
  should be done to make the extra dimension small and invisible,
  then the corresponding wedge-like sectors are cut out in the Kruskal diagram, hence
  the T-region and R-region in the $(r,v)$ subspace have only a single common point, 
  i.e., the horizon intersection point, as shown in Figure \ref{fig1}.  Instead of a horizon, we 
  thus obtain a naked conical singularity at $r=2m$.

  Quite another picture is observed if $\eta_v=-1$. A joint study of the regions on
  different sides of the horizon might again be possible after a transition to coordinates in 
  which the metric is manifestly nonsingular at $r =2m$. Let us perform such a transition 
  for (\ref{TScw}) in a close vicinity of the would-be horizon $r = 2m$ bearing in mind that other 
  more complex cases can be treated in a similar manner:
\bearr                                    \label{Thor}
		r - 2 m = \frac{x^2+y^2}{8m}, \qq   v = 4m \arctan \Big(y/x\Big), 
\nnn
	      ds^2_2 (r,v) \approx   \frac{r - 2m}{2m} dv^2 + \frac{2m}{r-2m}dr^2 = dx^2 + dy^2.
\ear

  Thus the $(r,v)$ surface metric is locally flat near the location $r=2m$, which is now 
  transformed to the origin $x=y=0$, while the $v$ coordinate becomes a multiple of
  the polar angle.
  
 \vspace{-16pt}    

  This transformation might be conducted as a conformal mapping of the complex plane 
  with the analytic function $\log z,\ \ z=x+ {\rm i}y$, then $v$ is proportional to $\arg\,z$.
  Such an operation was performed in \cite{br79} for some special cylindrically symmetric 
  Einstein-Maxwell solutions, see also \cite{cy20}.

\begin{figure}[H]    
\centering     
\includegraphics[width=9cm]{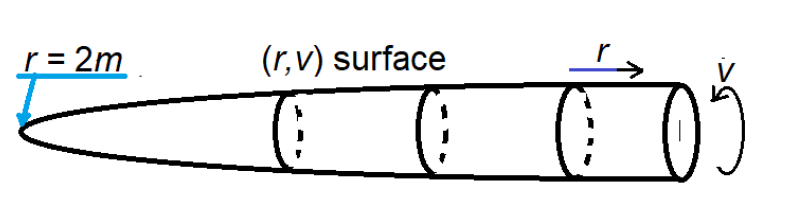}
\caption{\small
		A regular $r{-}v$ surface in a T-\Scw\ space-time with a spatial extra dimension.} 	
			\label{fig2}
\end{figure}

  As a result, in the general case, the $(r,v)$ surface near $r = 2m$ behaves like a
  Riemann surface for the function $\log z$ that possesses a finite or infinite (if $v$ varies 
  in an infinite range) number of sheets, with its branching point located at $x=y=0$, 
  which can be called a branch-point singularity, following \cite{br79}. Furthermore, 
  if the $v$ direction is compact, then, since $v$ now behaves as an angular coordinate,
  we can assume $0\leq v < 2\pi \ell$, where $v=0$ and $v=2\pi \ell$ are identified, 
  and $\ell$ has the meaning of a compactification length in the asymptotic region 
  ($r \to \infty$), where $g_{55} \to 1$.  Thus $r = 2m$ is the center of symmetry in the 
  $(r,v)$ surface, and from \rf{Thor} it follows that $r = 2m$ is a regular center under the
  condition $\ell = 2m$, it is an $n$-fold branch point if $l = 2m/n$, and at arbitrary 
  $\ell$ there is a conical-type singularity. With a regular center, the $(r,v)$ surface has 
  the shape of a test tube that has a constant radius $\ell = 2m$ at large $r$,
  becomes narrower at smaller $r$ and smoothly ends at $r = 2m$ (Figure \ref{fig2}).
  
     \vspace{-10pt}     
  \vspace{-16pt} 

  It means that a radial geodesic, whose projection to the ($r,v$) surface hits the point 
  $r=2m$, passes through it and returns to larger values of $r$, though with another 
  value of $v$, thus leaving a particular 4D section of space-time. However, if, instead 
  of a classical point particle, we consider a quantum one, with a $v$-independent 
  wave function, i.e., if it uniformly fills the $v$ direction, then the particle certainly does 
  not disappear from an observer's sight and can look as if reflected from a mirror. Such 
  an assumption looks natural if the compactification length is sufficiently small, as 
  should be the case in the framework of the Kaluza-Klein paradigm of extra dimensions. 
  Possible $v$-dependent wave functions would belong to particles with extremely 
  high energies corresponding to masses $\gtrsim 1/\ell$ in 4D.   
  
  We can conclude that this would-be horizon with $\eta_v=-1$ looks observationally like a 
  mirror and can be called a {\it mirror surface}. However, as follows from the regularity  
  condition $\ell = 2m$, such a mirror star has a very small mass: if we assume 
  $\ell \sim 10^{-18}$ cm according to the instrumental constraints, we will obtain 
  $m \sim 10^{15}\, m_{\rm Planck} \sim 10^{10}$ g, a mass that, if it were a \bh, 
  would evaporate via Hawking radiation in about one second.
  Quantum field effects close to a mirror surface can also be significant and deserve 
  a separate study.  Anyway, the direct relationship between this stellar mass and the 
  compactification length is a challenge, and it is of great interest whether more complex
  models can if not avoid then at least weaken this constraint.
  
\section{5D Einstein-Maxwell Fields: Black Holes and Mirror Stars} \label{sec3}

  As a next step, let us consider possible 5D analogs of the \RN\ solution of GR, 
  introducing the Maxwell field with the Lagrangian $L = - F \equiv - F_{AB} F^{AB}$ 
  as the only source of gravity with the \ssph\ metric \rf{ds5}. It is clear that the 
  solutions to be discussed are special cases of those considered in \cite{kb95a, kb95b}
  and much more general multidimensional solutions of \cite{bbm97,bh5}; it 
  nevertheless seems reasonable to discuss these simpler ones for better transparency 
  of their expressions and properties.   
  
  Under the assumptions made, in addition to radial electric ($F_{01} = -F_{10}$) and 
  magnetic ($F_{23} = -F_{32}$) fields, one more kind of  \emag\ field is compatible with 
  the symmetry of \rf{ds5}, the one corresponding to the component $A_5(u)$ of the
  vector potential $A_B$, leading to nonzero components $F_{15} = - F_{51}$ of the
  \emag\ tensor $F_{AB}$. Since such a component does not single out any 
  direction in 3D space, it will be called ``quasiscalar.'' The Maxwell equations 
\beq
		\D_A \Big(\sqrt{|g|} F^{AB}\Big) =0, \qq
		 \sqrt{|g|} = \e^{\alpha + 2\beta + \gamma + \xi} \sin\theta,
\eeq  
  for $F^{01}$  and $F^{15}$ depending on $u$ only are easily solved with the
  corresponding charges as integration constants. For a magnetic field we have 
  the usual expression corresponding to the potential component 
  $A_3 = q_m \cos\theta$. As a result, it can be jointly written 
\beq
		\Big\{F_{01}F^{01}, \ F_{23}F^{23}, \  F_{15}F^{15} \Big\}
		= \Big\{ -q_e^2 \e^{-4\beta - 2\xi},\
				 q_m^2 \e^{-4\beta},\ 
			- \eta_v q_s^2 \e^{-4\beta - 2\gamma} \Big\},
\eeq  
  where $q_e, q_m, q_s$ are the electric, magnetic and quasiscalar charges, respectively.
    
  Considering the existence of only one of the three kinds of \emag\ fields, we can write 
  the following expressions for their stress-energy tensors in all three cases: 
\bear     \label{SET-e}
	{\rm Electric:}\qq  && T_A^B = q_e^2 \e^{-4\beta - 2\xi} \diag(1,1,-1,-1,-1),
\yy	     \label{SET-m}
	{\rm Magnetic:}\ \ \ \ && T_A^B = q_m^2 \e^{-4\beta} \diag(1,1,-1,-1,1),
\yy     \label{SET-s}
	{\rm Quasiscalar:} \ && T_A^B = \eta_v q_s^2 \e^{-4\beta- 2\gamma } \diag(-1,1,-1,-1,1).	
\ear  

  One can notice that the energy density of a quasiscalar field is negative if the extra
  dimension is timelike. 

  The nonzero components of the Ricci tensor  for the metric \rf{ds5} can be written
  as follows without fixing the radial coordinate $u$ (the prime denotes $d/du$): 
\bearr            \label{Ric-gen}
		R^0_0 = -\e^{-2\alpha}\Big[\gamma'' +\gamma'(2\beta'+\gamma'+\xi'-\alpha')\Big],
\nnn		
  		R^1_1 = - \e^{-2\alpha}\Big[2\beta'' + \gamma'' + \xi'' 
    			+ 2\beta'^2 + \gamma'^2 + \xi'^2 - \alpha'(2\beta' + \gamma' + \xi')\Big],
\nnn
  		R^2_2 = R^3_3 = \e^{-2\beta}
  				- \e^{-2\alpha}\Big[\beta'' + \beta'(2\beta'+\gamma'+\xi'-\alpha') \Big],
\nnn
		R^5_5 = - \e^{-2\alpha}\Big[\xi'' + \xi'(2\beta'+\gamma'+\xi'-\alpha') \Big],	
\ear  
  where the coordinates are numbered as $(t, u, \theta, \varphi, v) = (0,1,2,3,5)$.
  Note that putting $\xi = \const$, we obtain the Ricci tensor components $R\mN$ of
  the 4D section of \rf{ds5}.  It is also useful to present the Einstein tensor component 
  $G^1_1 = R^1_1 - \half R$ since the corresponding Einstein equation 
  	(the 5D Einstein equations have the form $G^A_B \equiv R^A_B- \half R\delta^A_B =-T^A_B$) 
  is first-order and is an integral  of the others which are second-order:
\beq
		G^1_1 = - \e^{-2\beta} + \e^{-2\alpha}
				\big(\beta'^2 + 2\beta'\gamma' + 2\beta'\xi'  + \gamma'\xi'\big).
\eeq   

  A further solution is best of all carried out using the harmonic radial  coordinate $u$
  defined by the condition 
\beq              \label{harm}
            \alpha = 2\beta + \gamma + \xi 
\eeq
   in which case the Ricci tensor components take an especially simple form, as is evident 
   in \rf{Ric-gen}. Thus, the relation $T^1_1 +T^2_2 =0$, valid for all three kinds of 
   \emag\ fields, implies  $R^0_0 + R^2_2 + R^5_5 =0$, which in the coordinates 
   \rf{harm} takes the form of the easily solved  Liouville equation
\beq          \label{def-s}
		\alpha'' - \beta''  = \e^{2\alpha-2\beta} \ \ \then \ \
		 \e^{\beta - \alpha} = s(k,u) \equiv
     		 \vars{ k^{-1}\sinh\,ku,  & k > 0; \\
     				       u, & k = 0;  \\
     			   k^{-1}\sin ku, & k < 0, }    
\eeq
   where $k=\const \in \R$, and one more integration constant is eliminated by 
   choosing the zero point of $u$. Consequently, with no loss of generality, one can assert that 
   the harmonic $u$ coordinate is defined at $u > 0$, and $u = 0$ corresponds to spatial infinity. 
   
   The further solution looks differently for the three tensors \rf{SET-e}, \rf{SET-m} and 
   \rf{SET-s}. In all three cases we will obtain the general solution and single out the 
   cases containing horizon-like surfaces, which occur in the limit $u \to \infty$.
   
\subsection{Electric Field}   

  From different linear combinations of the Einstein equations it is easy to obtain 
\bearr           \label{nu-e}
		\xi'' + \Half\gamma'' = 0 \ \then  \xi = - \Half(\gamma + Nu) + \xi_0,
\yyy           \label{gamma-e}
           \gamma'' = Q^2 \e^{2\gamma} \ \ \then 
           		\e^{2\gamma} = \frac 1 {Q^2 s^2(h, u+u_1)},
\ear  
  where $N, h, u_1, \xi_0$ are integration constants, $Q = \sqrt{4/3} q_e$, and 
  the function $ s(h, u+u_1)$ is defined in the same way as in \eqn{def-s}. Two of the 
  integration constants, $u_1$ and $\xi_0$, can be fixed by choosing the units along 
  the $t$ and $v$ axes, in accord with asymptotic flatness at $u =0$:
\beq               \label{fix-u1}
		s^2(h,u_1) = Q^{-2}, \qq \xi_0 =0.
\eeq 
 
   One more relation between the integration constants is obtained by substituting 
   the expressions  obtained to the Einstein equation $G^1_1 = - T^1_1$:
\beq                         \label{int-e}
		k^2 \sign k = \frac 34 h^2 \sign h + \frac 14 N^2.
\eeq   
 
   Summarizing, we obtain the metric in the form
\beq                           \label{ds-e}
	ds_5^2 = \e^{2\gamma}dt^2 
	- \frac{\e^{-\gamma + Nu}}{s^2(k,u)}\bigg[\frac{du^2}{s^2(k,u)} + d\Omega^2\bigg]
	+ \eta_v \e^{-\gamma - Nu} dv^2,			   
\eeq   
  with $\e^\gamma$ presented in  \rf{gamma-e}. The solution consists of a few branches
  with different analytical behaviors depending on the signs of the constants $k$ and $h$.
  
  It can be verified that a horizon within this solution is only possible in the case 
  $h = k = N > 0$, and then the metric is presented in the most transparent form 
  after the substitution $u \mapsto x$ such that
\bearr                           \label{u->x}
	  \e^{-2ku} = 1-\frac{2k}{x}, \qq   du = \frac{dx}{x^2(1-2k/x)},
	  \qq   \sinh ku = \frac{1-k/x}{\sqrt{1-2k/x}},
\nnn
	 \sinh[k(u+u_1] = \frac kQ \,\frac{1+p/x}{\sqrt{1-2k/x}}, \qq p = \sqrt{k^2 + Q^2} -k.		  
\ear    

   This results in the black hole metric, \asflat\ as $x\to \infty$,
\beq                             \label{bh-e}
		ds_5^2 = \frac{1-2k/x}{(1+p/x)^2}dt^2- \bigg(1+ \frac px\bigg) 
					\bigg[\frac{dx^2}{1-2k/x}  + x^2 d\Omega^2 - \eta_v dv^2 \bigg]
\eeq   
   with an event horizon at $x = 2k$ and a singularity at $x=0$. Unlike the \RN\ space-time, 
   this one contains no Cauchy horizon, and the whole region $x < 2k$ has the nature of a 
   Kantowski-Sachs cosmology that ends with a quasi-isotropic ``big crunch'' since both
   Kantowski-Sachs scale factors $\sqrt{|g_{00}|}$ and $\sqrt{|g_{22}|}$ behave at small $x$
   as $\sqrt x$, while the extra dimension blows up, $-g_{55} \sim 1/x$.  The Carter-Penrose
   diagram for the $(t, x)$ subspace is the same as for the \Scw\ space-time.
   
\subsection{Magnetic Field}
 
  Instead of  \rf{nu-e} and \rf{gamma-e},  linear combinations of the Einstein equations 
  now give  
\bearr                              \label{mu-m}
		\mu'' = Q^2 \e^{2\mu} \ \then 
           		\e^{2\mu} = \frac 1 {Q^2 s^2(h, u+u_1)},
           		 \  \   {\rm where} \ \ \ \mu = \gamma + \xi,     
\yyy	                           \label{nu-m}	
		\gamma'' = \xi''\ \then\  
		\gamma = \Half (\mu + Nu), \qq  \xi = \Half(\mu - Nu),
\ear  
  where $N, h, u_1$ are integration constants, now $Q = \sqrt{4/3} q_m$, and,
  as before, we have fixed the constants in such a way that $\gamma(0) = \xi(0) =0$ 
  in accord with asymptotic flatness at $u =0$, and in particular, the relation \rf{fix-u1}
  holds for $s(h,u_1)$.  Also, as with the electric field, the Einstein equation 
  $G^1_1 = - T^1_1$ leads to the relation \rf{int-e} between the integration constants
  (note that the notations are chosen in a suitable way to obtain this similarity). 
  The resulting metric has the form
\beq                           \label{ds-m}
	ds_5^2 = \e^{\mu + Nu}dt^2 
	- \frac{\e^{-2\mu}}{s^2(k,u)}\bigg[\frac{du^2}{s^2(k,u)} + d\Omega^2\bigg]
	+ \eta_v \e^{\mu - Nu} dv^2,			   
\eeq   
  with $\e^\mu$ presented in  \rf{mu-m}, and there are again a few branches
  of the solution depending on the signs of the constants $k$ and $h$.
  
  Metrics without naked singularities are obtained from \rf{ds-m} in the cases $k=h >0$ 
  and $N=\pm k$. Applying again the transformation \rf{u->x}, we obtain a 
  \bh\ metric in the case $N=-k$,
\beq		                           \label{bh-m}
		ds_{5, \rm bh}^2 = \frac{1-2k/x}{1+p/x} dt^2- \Big(1+ \frac px\Big)^{\!2}\ 
					\bigg[\frac{dx^2}{1-2k/x}  + x^2 d\Omega^2\bigg] 
					+ \frac {\eta_v dv^2}{1 + p/x},
\eeq  
   and a mirror star metric if $N = k$:
\beq		                           \label{mms}
		ds_{5, \rm ms}^2 = \frac{dt^2}{1+p/x} - \bigg(1+ \frac px\bigg)^{\!2}\  
					\bigg[\frac{dx^2}{1-2k/x}  + x^2 d\Omega^2\bigg] 
					+ \frac {1-2k/x}{1 + p/x} \eta_v dv^2,
\eeq     
   and they are related, as noted in the Introduction, by the replacement
   $dt^2 \rightleftarrows \eta_v dv^2$. The constants used here, $p > 0$ and $k >0$, 
   are related by  
 \beq      \label{k->q}
 		p = \sqrt{k^2 + Q^2} -k \ \then \ k = \frac{4q^2 - 3p^2}{6p}, 
 \eeq   
   where as before, $Q^2 = (4/3) q^2$. Each of these two metrics contains two 
   independent physical parameters, for which one can choose the mass $m > 0$ 
   and the charge $q= q_m$, and both metrics are conveniently rewritten in terms of 
   the spherical radius $r = x+p$; however, they describe drastically different geometries,
   so let us discuss them separately. 
   
\paragraph{Black holes:} with $r = x+p$, the metric \rf{bh-m} takes the form
\beq   	                           \label{mbh-r}
		ds_{5, \rm bh}^2 = \frac{r-2m}{r} dt^2 - \frac {r^2 dr^2}{(r-2m)(r-p)} 
			- r^2 d\Omega^2 + \frac{r-p}{r} \eta_v dv^2,
\eeq   
   with the \Scw\ mass $m =(p + 2k)/2$. Since $k > 0$, we have $p < 2m$, and with  
   \rf{k->q} it follows
\beq              \label{qm-bh}
		3 mp = 2q^2, \qq    q^2 < 3 m^2.
\eeq   

   This space-time contains an event horizon at 
   $r = 2m >p$ and one more peculiar sphere $r=p>0$, inside which the 4D metric acquires 
   the Euclidean signature. Furthermore, if the extra coordinate $v$ is temporal outside 
   the object ($\eta_v = +1$), then the whole 5D metric becomes Euclidean at $r < p$ 
   since all five coordinates become spatial, and the cosmological Kantowski-Sachs 
   evolution terminates at $r=p$. Unlike that, if $\eta_v = -1$, then $r =p$ is a conventional
   Killing horizon in the $(r,v)$ subspace. At $r \in (p, 2m)$, the only temporal coordinate 
   is $r$, while at $r < p$ the only temporal coordinate is $v$. And this time variable is 
   circular as long as $\{v\}$ is a circle. In all cases, $r=0$ is a curvature singularity. What is 
   important, in such \bh\ solutions the compactification length $\ell$ is not related 
   to the mass $m$ and charge $q$, and all these parameters are arbitrary (up to \rf{qm-bh}).  

\paragraph{Mirror stars:} from \rf{mms}, in terms of $r = x+p$ we obtain
\beq		                           \label{mms-r}
		ds_5^2 = \frac{r-2m}{r} dt^2 - \frac {r^2 dr^2}{(r-2m)(r-r_b)} - r^2 d\Omega^2
				+ \frac{r -r_b}{r} \eta_v dv^2.
\eeq  
  
    Here the \Scw\ mass is $m = p/2$, but the sphere $r=2m$, corresponding to the 
    observable mass, is not an event horizon since a larger radius, $r_b = 2m + 2k$ is that 
    of a mirror surface, which represents a boundary of this space-time as described above.     
    Now, with \rf{k->q}, we obtain, instead of \rf{qm-bh},
\beq    		              \label{qm-ms}
		3 m r_b = 2 q^2, \qq   q^2 > 3 m^2,
\eeq    
    and at smaller $q^2$ this solution does not exist.
    In the metric \rf{mms-r}, the expression of  $g_{tt}$  does not reach zero as $r$ 
    decreases: the solution terminates at larger $r$, so that a mirror star is less compact 
    than a would-be \bh\ of the same mass.
   
   A few words on the geodesic structure in this space-time. In any \ssph\ space-time
   with the 4D part of the metric \rf{ds5}, 4D geodesics (those with $v=\const$) satisfy the 
   equation (see, e.g., \cite{br-book})
\beq                  \label{geo}
		\e^{2\alpha+2\gamma} \bigg(\frac{du}{d\lambda}\bigg)^2 + V(u) = E^2,
	\qq
		V(u) = \e^{2\gamma} (K + L^2/r^2),\qq   K =0, \pm 1,
\eeq   	
   where $u$ is an arbitrary radial coordinate, $L$ and $E$ are constants responsible for 
   the conserved angular momentum and energy of a particle, $\lambda$ is an affine 
   parameter along timelike ($K=1$), spacelike  ($K=-1$) and null  ($K=0$) geodesics,
   respectively, and $V(u)$ is the effective potential, whose properties actually determine
   all features of the geodesic motion. Now, since $g_{00}=\e^{2\gamma}$ and
   $g_{22} = r^2$ in \rf{mms-r} are the same as in the \Scw\ space-time, the 4D geodesic 
   structure is also the same outside the boundary $r = r_b  $. In particular, if $r_b < 3m$,
   then $r =3m$ is an unstable photon sphere, while the boundary itself is a stable photon
   sphere.   
   
   If we require regularity of the metric at $r = r_b  $, then a small size $\ell$ of the 
   fifth dimension strongly restricts the solution parameters, similarly to Section 2:
\beq
		4(m+k)^3 = k\ell^2. 
\eeq  
   
   \textls[-15]{The metric \rf{mms-r} coincides with the so-called ``topological star'' metric discussed in 
   \cite{tops1, tops2, tops3} under the identification }
\beq
		   r_b   = 2m +2k \equiv r_B, 	\qq  r_{\rm Scw} = 2m \equiv r_S.
\eeq 
 
   Thus $r_B$ is the radius of the mirror sphere, while $r_S$ is the \Scw\ radius for the same mass. 
   
{
\paragraph{Intermediate case: extremal \bhs} At $q^2 = 3m^2$, that is, $p =2m$ 
   in \rf{mbh-r}, or equivalently $r_b = 2m$ in \rf{mms-r}, we obtain the metric
\beq				                           \label{ex-bh}
		ds_5^2 = \frac{r - 2m}{r} dt^2 - \frac {r^2 dr^2}{(r-2m)^2} - r^2 d\Omega^2
				+ \frac{r - 2m}{r} \eta_v dv^2.
\eeq 

   That it describes an extremal \bh\ is made evident by using, for example, the new coordinate 
   $x$ defined by $r = 2m + x^2$, so that the metric reads
\beq				                           \label{ex-bh-x}
		ds_5^2 = \frac{x^2}{(2m + x^2)^2} dt^2 - \frac {4 (2m + x^2)^2 dx^2}{x^2} 
				- (2m + x^2)^2 d\Omega^2+ \frac{x^2}{(2m + x^2)^2} \eta_v dv^2.
\eeq 

   In this metric, $x \in \R$, and $x < 0$ describes a region beyond the horizon $x=0$, with 
   a geometry identical to $x > 0$. Curiously, in this case spherical radii smaller than $2m$
   are impossible, and the whole system resembles a \wh, but it is not traversable due to
   a horizon at its throat.    }
    
{A separate solution may be obtained in the same manner for $r \leq 2m$, but it seems
   to be of lesser interest and will not be discussed here.}    
   
\subsection{Quasiscalar Fields}

   The results in this case are similar to those for an electric field with the substitutions
   $t\chg v$ and $\gamma \chg \xi$ but taking into account the two signs in 
   $\eta_v = \pm 1$. Thus, instead of \rf{nu-e} and \rf{gamma-e} we now obtain
\bearr                         \label{nu-s}
             \xi'' = \eta_v \e^{2\xi} \ \then  \e^{-2\xi} 
             			=  \vars{ s^2 (h, u+u_1),  &  \eta_v = +1, \yy
     				      		        h^{-2} \cosh^2 [h(u+u_1)],\ h>0, & \eta_v = -1, } 
\yyy                          \label{gamma-s}
		\gamma'' + \Half\xi'' = 0 \ \then  \gamma = - \Half(\xi + Nu),
\ear   
   with integration constants $N, h, u_1$ and $Q = \sqrt{4/3} q_s$ and again adjusting to 
   $\gamma(0) = \xi(0) =0$. As before, \eqn{int-e} is valid. The resulting metric has the form
\beq                            \label{ds-s}
               ds_5^2 = \e^{-\xi -Nu} dt^2 
            	- \frac{\e^{-\xi + Nu}}{s^2(k,u)}\bigg[\frac{du^2}{s^2(k,u)} + d\Omega^2\bigg]
		+ \eta_v \e^{2\xi} dv^2,		
\eeq  
    where $\e^\xi$ is given by \rf{nu-s}, and there are even more branches of the 
    solution with different $k, h$ and $\eta_v$.
    
    As before, solutions without naked singularities emerge in the case $k = h = N >0$, which, 
    after the substitution \rf{u->x}, leads to the metric
\bearr
		ds_5^2 = \Big(1 + \frac px\Big) \bigg[ dt^2 
		                - \frac {dx^2}{1- 2k/x} - x^2 d\Omega^2 \bigg] + \eta_v \e^{2\xi} dv^2,
\nnn                                              \label{s+}
		\e^{2\xi} = \frac{1-2k/x}{(1+p/x)^2}, \qq \eta_v =1, 		
\nnn                                       \label{s-}
		\e^{2\xi} = \frac{Q^2(1-2k/x)}{(k + p - kp/x)^2}, \qq  \eta_v = -1. 		                
\ear

    As follows from the previous consideration, only in the case \rf{s-} this metric can 
    correspond to a mirror star. Furthermore, as follows from the expression of $g_{tt}$, 
    since $p > 0$, we hear deal with a negative \Scw\ mass $m = -p/2$, therefore 
    this metric does not seem to be physically relevant.  
     
    Thus, among the electrovacuum solutions, the magnetic solutions with the metrics 
    \rf{bh-m} and \rf{mms} are of particular interest, and in what follows we will discuss
    their stability.              
       
\section{Magnetic Solutions: Reduction to 4D and the Stability Problem}   \label{sec-stab} 
\subsection{General Consideration}

  Now we would like to determine the stability properties of the magnetic mirror star solution 
  \rf{mms}. While the solution itself was obtained directly in the 5D setting, a stability study 
  turns out to be easier in the 4D setting using the Einstein conformal frame. 
  
  Thus, we are considering the 5D action with the matter Lagrangian $L_m$
\beq        \label{S_5}
		S = \int d^5 x \sqrt{^5g}  (^5 R + L_m) 
\eeq  
   in a manifold with the 5D metric  
\beq        \label{ds_5g}
		ds_5^2 = g\mn dx^\mu dx^\nu + \eta_v \e^{2\xi(x)} dv^2,  
 \eeq
  where, as before, $\eta_v = \pm 1$, the 4D metric $g\mn$ and the metric coefficient
  $g_{55} =\e^{2\xi(x)} $ do not depend on the 5th coordinate $v$. Then, integrating it 
  out, we obtain, up to a constant factor and a full divergence,
\beq        \label{S-4J}
		S =  \int d^4 x \sqrt{^4 g} \e^{\xi(x)} (^4 R + L_m).     
\eeq

  The next step is to transform it to the Einstein frame (E-frame) by the standard conformal mapping
\beq         \label{conf}
		g\mn \to \og\mn = \e^{\xi} g\mn,
\eeq		
  which converts \rf{S-4J} to the form (again up to a full divergence)
\beq        \label{S4-E}
		S =  \int d^4 x \sqrt{^4 \og} 
			\Big({}^4\oR + \frac 32 \og\MN \xi_\mu \xi_\nu + e^{-\xi} L_m\Big), 
\eeq   
   where $\xi_\mu:= \D \xi/\D x^\mu$, and the overbar marks quantities obtained 
   from or with the metric $\og\mn$. Thus the 5D equations are reduced to 4D 
   equations with a massless scalar field $\xi$ minimally coupled to gravity but 
   interacting with matter. The matter Lagrangian itself, in general, depends on 
   the 5D metric, which is now reduced to the quantities $\og\mn$ and $\xi$. Since 
   we are considering electrovacuum solutions, let us look how  the electromagnetic 
   invariant $\cF = F_{AB}F^{AB}$ will appear in the action \rf{S4-E}.
   
   If there are only 4D components $F\mn$ while $F_{5\mu} =0$, we have 
   $\cF = F_{AB}F^{AB} = F\mn F\MN$, and, since the expression  $\sqrt{|g|} F\mn F\MN$ 
   is invariant under conformal transformations, the same expression is valid in terms of the 
   metric $\og\mn$.  In particular, if $L_m = -\cF$ (the 5D Maxwell Lagrangian), then 
   \eqn{S4-E} takes the form well known as the action of dilaton gravity. 
   
   Unlike that, if $F_{5\mu} =\D_\mu \chi \equiv \chi_\mu\ne 0$, where 
   $\chi (x^\mu)= A_5$ is the extra-dimensional component of the vector potential
   (a quasiscalar field, see Sec.\,3), we obtain 
   $F_{5\mu} F^{5\mu} = \eta_v \e^{-2\xi} \og\MN \chi_\mu \chi_\nu$. Assuming 
   that $F\mn =0$ (that is, $F_{5\mu}$ are the only nonzero components of $F_{AB}$), we find 
\beq                  
		 \cF = F_{AB}F^{AB} = 2 \eta_v \e^{-2\xi} \og\MN \chi_\mu \chi_\nu.
\eeq   

  Thus we obtain an effective scalar field $\chi$ in 4D, this field is canonical if  
  $\eta_v = -1$ (a spacelike extra dimension) and phantom if $\eta_v =1$. As a result,
  in this case we have in 4D a special case of a sigma model with two interacting scalars
  $\xi$ and $\chi$, and its consideration is beyond the scope of this study.
  
  So let us return to the action \rf{S4-E} with $L_m = - F\mn F\MN$ and write the E-frame 
  metric in the form
\beq
		\overline{ds}{}^2 = 
			 \e^{2\ogamma} dt^2 - \e^{2\oalpha} dx^2 - \e^{2\obeta} d\Omega^2,
\eeq  
   where the barred notations $\oalpha, \obeta, \ogamma$ are similar to 
   $\alpha, \beta, \gamma$ used in \rf{ds5} but refer to our 4D E-frame metric.
   
   For \sph\ \pbs\ of such models it is well known that they are determined by a single degree
   of freedom associated with the effective scalar field $\xi$ of extra-dimensional origin,
   and since the action \rf{S4-E} is a special case of the action (A1) in \cite{GC15} 
   (with $\phi \mapsto \xi, h = 3/4, V=0, S(\phi)\mapsto \e^\xi, q_e=0$), we can actually use the 
   \pb\ equations obtained there. 
   In this situation, the metric \pbs\ $\delta\oalpha, \delta\obeta, \delta\ogamma$
   are also nonzero, but they only exist due to the \sph\ scalar \pb\ $\dxi$, being connected
   via the Einstein equations. 
   
  Let us show this in more detail. Suppose that our equations due to \rf{S4-E} are 
  solved for the static, time-independent variables $\xi, F\mn, \oalpha, \obeta, \ogamma$.
  Moreover, we are dealing with a magnetic solution, so that all $F\mn$ other than 
  $F_{23}=-F_{32} = q\sin \theta$ are zero, and $F\mn F\MN = 2q^2 \e^{-4\obeta}$;
  note that this remains true in perturbed space-time. Then the scalar field equation 
  for small time-dependent \pbs\ (``deltas'') can be written as follows:
\beq   			\label{wave-i}
		-\e^{2\oalpha-2\ogamma}\delta\ddot{\xi} + \dxi''
		 		+ (2\obeta'+\ogamma'-\oalpha') \dxi' + 
		 		    \xi' (2\delta\obeta' + \delta\ogamma' - \delta\oalpha') 
		 		    - \frac 23 q^2 \delta \Big(\e^{2\oalpha - 4\obeta + \xi}\Big) = 0,
\eeq
  where dots and primes denote $\D/\D t$ and $\D/\D x$, respectively,
  and the quantities $\xi, \oalpha, \obeta, \ogamma$ are taken from the static (``background'') 
  solution. Now, the Einstein equations make it possible to express the metric \pbs\ in terms of $\dxi$.
  In doing that, a necessary step is to fix the \pb\ gauge, which means making a particular choice
  of a reference frame in perturbed space-time, and this is implemented by postulating some 
  relation for the \pbs. 
  
{The relevant components of the Einstein equations (used in the form
   $R\mN = - (T\mN - \half \delta\mN T^\rho_\rho)$) are $(tx)$ and $(\theta\theta)$. Without 
   choosing a \pb\ gauge they read, respectively,   
\bearr   
   2[\delta\dot\obeta' + \obeta' \delta\dot{\obeta}                        \label{R01}
                 - \obeta'\delta\dot{\oalpha} - \ogamma' \delta\dot{\obeta}] =  - \frac 32 \xi' \delta{\dot \xi}, 
\yyy           \label{R22}
   \delta(\e^{-2\oalpha-2\obeta}) +\e^{2\oalpha-2\ogamma}\delta\ddot\obeta
              - \delta[\obeta''+ \obeta'(\ogamma'-\oalpha'+2\obeta')]
		              		              = 	q^2 \delta\big(\e^{2\oalpha - 4\obeta +\xi}\big).
\ear 
 }
 
  It turns out that the gauge $\delta\obeta \equiv 0$ substantially simplifies the equations, so in 
  what follows we employ this gauge. Then the $(tx)$ component of the Einstein equations can be 
  used to express $\delta\oalpha$ in terms of $\dxi$ and functions from the background solution: 
\beq            \label{d-al}
		2 \obeta' \delta\dot\oalpha  = \frac 32 \xi' \delta{\dot \xi} 
		\ \ \then \ \
		\delta \oalpha = \frac {3 \xi' \dxi}{4\obeta'}.
\eeq   

   Here, performing integration in $t$, we omit adding an arbitrary function of $x$ since we
   are only interested in time-dependent \pbs. 
   Next, the component $(\theta\theta)$ of the Einstein equations expresses the difference 
   $\delta\oalpha'- \delta\ogamma'$ in terms of $\delta \oalpha$ and $\dxi$. Specifically, 
\beq   \label{d-ga-al}
		\delta\ogamma' - \delta\oalpha'
				= \frac{1}{\beta'}\Big[\delta\Big(\e^{2\oalpha - 2\obeta}\Big)
									- q^2 \delta\Big(\e^{2\oalpha - 4\obeta +\xi}\Big) \Big].
\eeq  
 
   Substituting \rf{d-al} and \rf{d-ga-al} into the scalar wave equation \rf{wave-i}, 
   we exclude from it all metric \pbs, with the result
\beq   			\label{wave-0}
		-\e^{2\oalpha-2\ogamma}\delta\ddot{\xi} + \dxi''
		 		+ (2\obeta'+\ogamma'-\oalpha') \dxi' - U(x) \dxi =0,
\eeq   
   where
\beq   			\label{U(x)}
		U(x) = \e^{2\oalpha} \bigg[
			\frac{3\xi'^2}{2\obeta'^2}\Big(q^2\e^{\xi-4\obeta} - \e^{-2\obeta}\Big)
			+ q^2\e^{\xi-4\obeta} \bigg(\frac{2\xi'}{\obeta'} + \frac 23\bigg)	 \bigg],
\eeq
   and, which is important, the choice of the radial coordinate is so far not fixed. Let us use 
   this freedom, passing on to the ``tortoise'' coordinate $z$ (such that $\og_{tt}= -\og_{zz}$),
   and also replace the unknown function, $\dxi \to \psi$, according to the relations
\beq                 \label{xz}
       \frac {dx}{dz} = \e^{\ogamma - \oalpha}, \qq
   	\dxi (x,t) = \e^{-\obeta} \psi(x,t).
\eeq

  Then we arrive at the wave equation in its canonical form,
\beq  			\label{wave}
		- \ddot\psi  + \frac {d^2\psi}{dz^2} - \Veff(z) \psi=0,
\eeq
   with the effective potential 
\beq 			\label{Veff}
		\Veff(z) = \e^{2\ogamma - 2\oalpha}
				\Big[U + \obeta'' + \obeta'(\obeta'+\ogamma'-\oalpha')\Big].
\eeq

     The standard substitution $\psi(z) = \e^{i\omega t} Y(z)$ with some constant
   ``frequency'' $\omega$ converts \rf{wave} to the \Schr-like form
\beq  			\label{Schr}
		\frac {d^2 Y}{dz^2} + \big[\omega^2 - \Veff(z)\big] Y=0.
\eeq

   Equation \rf{Schr} with proper boundary conditions allows for a stability study, 
   such that eigenvalues $\omega^2 \leq 0$ correspond to \pbs\ growing with time.
   
   It should be stressed that \eqn{Schr} and, in particular, the potential $\Veff$ are
   gauge-invariant, that is, are independent from the choice of a reference frame in 
   perturbed space-time, and hence describe genuine physical \pbs. This invariance was
   explicitly proved in \cite{BFZ11} by directly considering gauge transformations and 
   is reflected in the fact that $\Veff$ depends only on the functions characterizing the
   background solution.
   
   We see that \pbs\ of the effective scalar field $\xi$ of \mult\ origin determine
   all metric \pbs: specifically, in our gauge, $\delta\obeta =0$,  $\delta\oalpha$ is 
   expressed in terms of $\dxi$ in \eqn{d-al}, and $\delta\ogamma'$ (using known 
   $\delta\oalpha$) can be found from \eqn{d-ga-al}; this determines $\delta\ogamma$
   up to adding an arbitrary function of $t$ that reflects arbitrariness of the time coordinate
   in a nonstatic space-time. In their turn, \pbs\ of the original 5D metric \rf{ds5} are found in 
   terms of $\dxi$ according to the transformation \rf{conf}:
\beq                \label{dg_mn}
		\delta\alpha = \delta\oalpha - \dxi/2,\qq 
		\delta\beta = -\dxi/2, \qq
		\delta\gamma = \delta\ogamma - \dxi/2.
\eeq   
      
\subsection{Spherical Perturbations of Magnetic Mirror Stars}

  As said above, this solution depends on two physical parameters, the mass $m$ and 
  the charge $q$, but it turns out to be more convenient to use, instead of $q$, the mirror 
  horizon, or boundary, radius $r_b = 2q^2/(3m)$.
   
  The scalar $\xi$ and the components of the E-frame metric 
  $\og\mn$ for the mirror star solution are given by 
\bearr          \label{ds-ms}
		\e^{\xi(x)} = \frac{\sqrt{r-r_b  }}{\sqrt{r}},
	\qq
		\e^{2\ogamma} = \e^{-2\oalpha} = \frac{(r-2m)\sqrt{r-r_b  }}{r^{3/2}},
	\qq
		\e^{2\obeta} =\sqrt{r-r_b  }r^{3/2},			  
\ear

\begin{figure}[H]    
\centering     
\includegraphics[width=7.2cm]{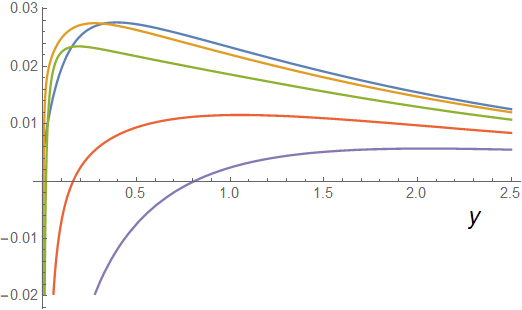}
\includegraphics[width=8cm]{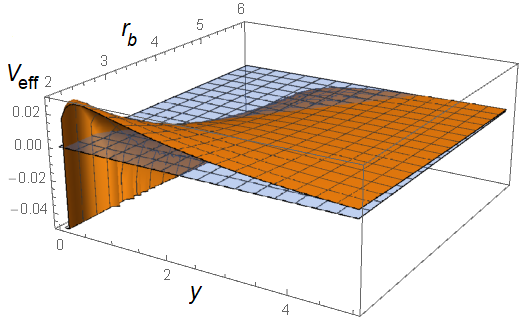}
\caption{\small
		(\textbf{Left}): The effective potential $\Veff$ for mirror star \pbs\
           as a function of $y = r-r_b  $ for $m=1$ 
		and $r_b -2m= 0.02, 0.1, 0.3, 0.7, 1.3$ (upside-down at larger $y$). 
		(\textbf{Right}): 3D plot of $\Veff(y)$ for $m=1$ and $r_b  - 2m \in (0,4)$.		
		The transparent level $\Veff =0$ visualizes the region where $\Veff >0$.
		Note that the charge $q$ is given by $q^2 = (3/2) m r_b  $.
		}     \label{fig-V}
\end{figure}

   Accordingly, for the metric \rf{ds-ms} we obtain 
\beq   			\label{Ums}
		U(x) = - \frac{2 r_b   (3 r_b -8 m)}
				{(r - 2 m) (4 r - 3 r_b  )^2 (r - r_b  )}.
\eeq    
  and the effective potential
\bearr 			\label{Veff-ms}
	\Veff(x) = -\frac{r - 2m}{16 r^5 (4 r - 3 r_b  )^2 (r - r_b  )}\ 
      \Big[(512 m r^4 + 128 r^3 (-13 m + r) r_b 
\nnn \inch      
      + 16 (152 m - 27 r) r^2 r_b^2 + 24 r (-69 m + 16 r) r_b^3 + 27 (14 m - 3 r) r_b^4)\Big],
\ear   
   its behavior is illustrated in Figure \ref{fig-V}.

   At large $r$ and near the surface $r = r_b  $, the potential \rf{Veff-ms} looks approximately as
\beq 		\label{as-Veff-x}
		\Veff(x)\Big|_{r\to\infty} = \frac{4m + r_b  }{2r^3} + O(r^{-4}),
		\qq
		\Veff(x)\Big|_{r\to r_b  } = -\frac{(r_b  -2m)^2}{16 r_b  ^2 (r - r_b  )} + O(1).
\eeq 
  
   The tortoise coordinate $z$, related to any $x$ by \rf{xz} (now with $dx= dr$), has the 
   asymptotic behavior
\beq 		\label{as-z} 
		z \approx r \ \ {\rm as}\ \  r\to \infty, \qq 
		z\approx \frac{2 r_b  ^{3/2}\sqrt{r-r_b  }}{r_b  -2m}	 \ \ {\rm as}\ \  r\to r_b  ,
\eeq   
   assuming $z=0$ at $r=r_b$ without loss of generality. Therefore, in terms of $z$,
\beq 		\label{as-Veff-z}
		\Veff(z)\Big|_{z\to\infty} = \frac{4m + r_b  }{2z^3} + O(z^{-4}), 
		\qq
		\Veff(z)\Big|_{z\to 0} = -\frac{1}{4 z^2} + O(1).
\eeq   

\vspace{-6pt}
 \vspace{-16pt}

   A study of the properties of $\Veff$ allows for qualitative inferences. Thus, using \rf{as-Veff-z}, 
   it is easy to obtain the asymptotic form of solutions to \eqn{Schr} at large and small $z$ for 
   fixed $\omega^2 < 0$ (here and further on all $C_i = \const$):
\bearr			\label{Y-inf}
		z\to \infty:\qq      Y(z) = C_1 \e^{i\omega z} + C_2 \e^{-i\omega z},
\yyy			\label{Y-2k}
		z\to 0: \qq\       Y(z) = \sqrt{z}(C_3 + C_4 \log z),
\ear    

    Now, let us discuss the boundary conditions. 
{At large $z \approx r$, a natural requirement is a finite total energy of \pbs. The scalar 
    field energy density is $\rho_\xi \sim \xi'^2$, so we have in the linear approximation 
    $\delta\rho_\xi \sim \xi'\dxi'$, and since according to \rf{ds-ms} $\xi' \sim 1/r^2$ at large $r$, 
    for finiteness of the total energy we require $\rho_\xi = o(1/r^3)$, hence $\dxi' = o (1/r)$.
    On the other hand, if $\im\omega \ne 0$, one of the terms in \eqn{Y-inf} exponentially 
    vanishes at large $z$ while the other blows up, the latter being evidently incompatible with 
    $\dxi' \sim (Y/r)' = o (1/r)$. Therefore, only the exponentially vanishing term is suitable, and
    our boundary condition for $Y$ at large $r$ must read simply $Y \to 0$.}

    Other subtle points must be discussed for the boundary $z=0$. 
    In similar cases, where $z=0$ is a singularity and the effective potential behaves in the 
    same way (see \cite{kb-hod, we23} among others), with the background scalar field 
    $\xi \sim \log (r-r_b) \sim \log z$, and $Y \sim \psi \sim \dxi \sqrt{z}$, there seems to 
    be no reason to require that $\dxi$ should blow up (if at all) slower than $\xi$ itself.
    Then admitting $\dxi \sim \xi$ at small $z$, we obtain the boundary condition 
    $Y \lesssim \sqrt{z} \log z$, so that all solution to \eqn{Schr}, with any finite 
    $\omega$, satisfy this condition. Hence physically admissible are \pbs\
    growing in time with any increment $|\im\omega|$, which means that the background 
    configuration is catastrophically unstable, or, in other words, even small perturbations, 
    having emerged, immediately show a nonlinear behavior.
    
    However, in the present case, the value $z=0$ corresponds to a regular surface in 5D 
    space-time, and it is reasonable to require that \pbs\ preserve this regularity even though 
    $g_{55} = -\e^{2\xi}$ vanishes there. It can be shown that the regularity requirement 
    leads to $Y \lesssim \sqrt{z} \ \then\ C_4 =0$. This can be most easily verified using the 
    trace of the 5D Einstein equation that gives   
\beq    
		    R = \frac 23 T = \frac 23 q^2 \e^{-4\beta} = \frac 23 q^2 \e^{-4\obeta + 2\xi}
\eeq    
    (taking into account \rf{dg_mn}). Since $\delta\obeta=0$, a \pb\ of the 5D curvature 
    scalar $R$ is {simply $\delta R = 2 R\dxi$ while $R(r_b) = \frac 23 q^2 r_b^{-4} >0$}, 
    and to keep $R$ finite we must require $|\dxi| < \infty$. Recalling that 
    $\dxi \sim Y/\sqrt z$ near $z=0$ ($r = r_b$), we come to the condition $Y/\sqrt{z} < \infty$.

    We conclude that our stability study requires solving a boundary-value problem for
    \eqn{Schr} with the boundary conditions
\beq           \label{BC-z}
			Y \to 0  \ \ {\rm as}\ \  z\to \infty,
			\qq
			Y/\sqrt{z} < \infty  \ \ {\rm as}\ \  z\to 0.
\eeq     
   
{It is then straightforward to show that the \Schr\ operator $-d^2/dz^2 + \Veff(z)$ with 
  the boundary conditions \rf{BC-z} is self-adjoint in the Hilbert space of square-integrable
  functions $Y(z)$, which, according to the Sturm-Liouville theory, ensures that its 
  spectrum is purely real, $\omega^2\in \R$. Therefore, in our search for unstable modes
  of \pbs, we must seek solutions to \eqn{Schr} with $\omega^2 <0$, hence pure imaginary 
  $\omega$, and $Y(z) \sim\e^{-|\omega| z}$ as $z\to \infty$.}

{Let us stress that in the present approach we study \pbs\ emerging in our
  ``isolated'' system itself, rather than due to any signal or pumping coming from outside. 
  This approach is, however, different from the one used in the analysis of quasinormal 
  modes \cite{bh4, qnm09}, where one requires that only outgoing waves are present, and 
  one considers a field $\Phi \sim \e^{-i\omega(t-z)}$ at $\im\omega <0$ corresponding to a 
  signal decreasing in time but growing at large $z$. Nevertheless, we can notice that
  in our case unstable \pbs\ growing exponentially in time and decaying at infinity
  also correspond to outgoing waves at large $z$. }
    
  Now, passing on to a numerical study, we have to take into account that the potential      
  $\Veff$ is expressed in terms of the coordinate $r$, connected with $z$ by a transcendental 
  equation, therefore, such a study is more conveniently conducted using
  \eqn{wave-0} written in terms of $r$. The substitution $\dxi(x) = \e^{i\omega t} X(x)$
  (such that $X(r) =  \e^{-\obeta} Y(z) \approx Y(z)/\sqrt{z}$) brings \eqn{wave-0} to the form 
 \beq   			\label{X''}
		X'' + (2\obeta'+\ogamma'-\oalpha') X'
					+\big(\e^{2\oalpha-2\ogamma}\omega^2 - U(r)\big) X =0,
\eeq   
  and the boundary conditions \rf{BC-z} are equivalently rewritten as
\beq                          \label{BC-x}
			X \toas_{r\to \infty} 0, \qq |X(r_b)| < \infty.		
\eeq   
     
 \subsection{Numerical Analysis for Magnetic Mirror Stars}  

  An attempt to numerically solve \eqn{X''} by specifying the boundary condition at large 
  $r$ using \eqn{Y-inf} with $C_1=0$ leads to a numerical instability near $r = r_b  $. 
  Therefore, let us try to use a ``left to right'' shooting procedure. A boundary precisely at
  $r = r_b  $ cannot be specified as it is a singular point of the equation. So let us choose a 
  boundary condition close to $r_b$,
  where a desired solution to \eqn{X''} should tend 
  to a constant value. Suitable boundary conditions can be chosen by finding the 
  asymptotic form of the solution $X(r)$ near $r = r_b$ in the next approximation to 
  $X = X_0 = \const$ (the assumed finite value at $r = r_b$).
  
  For convenience, let us rewrite \eqn{X''} close to $r = r_b$ in terms of $y = r - r_b$. It reads
\beq                \label{X''y}
		X'' + \frac{X'}{y} + \frac Ky X =0, \qq 
		{K = \frac{r_b^3}{(r_b-2m)^2}\omega^2 + \frac{6r_b - 16m}{r_b  (r_b - 2m)}.} 
		\eeq
		
   In general, it can be reduced to the Bessel equation, but we need its solution 
   only close to $y=0$ and seek it in the form $X = X_0 + c y^s,\ s>0$. We find
\beq
		s =1, \qq   X = X_0(1 - K y).
\eeq

  Therefore, a test function $X(y)$ in the shooting procedure with \eqn{X''} 
  should satisfy the following  boundary conditions at some small $y = y_0$:
\beq                      \label{BC-y}
             X(y_0) = X_0 (1 - K y_0), \qq   X'(y_0) = -K X_0,
\eeq   
  with $K$ given in \rf{X''y}. The equation itself in terms of $y$ has the form
\bearr
		X'' + \frac{2y +r_b   - 2m}{y (y+r_b  -2m)} X' 
				+ \bigg[\frac{(y+r_b  )^3}{y(y+r_b  -2m)^2}\omega^2  - U(y)\bigg] X =0,
\nnn
		{U = - \frac{2r_b(3r_b-8m)}{y(y+r_b-2m)(4y+r_b)^2}}
\ear  

   The goal is to find such $\omega^2$ that the solution $X(y)$ tends to zero as $y\to \infty$.

   To implement the shooting method, we use the standard Runge-Kutta  procedure for 
   solving the Cauchy problem \eqref{X''y}, \eqref{BC-y}, where $y$ belongs to the interval 
   $(y_0, y_1)\sim(10^{-3}, 10^3)$ giving the appropriate numerical accuracy for our purposes, 
   and $\omega^2$ is regarded as an unknown parameter varying inside some range 
   $({\omega^2_{\min}, 0})$. For each test value of $\omega^2$, we obtain the corresponding
   numerical curve $X_{\rm num}(y; \omega)$ satisfying the initial conditions \eqref{BC-y}. 
   At the right end $y_1$, the value of $X_{\rm num}(y; \omega)$ strongly diverges if 
   $\omega^2$ does not coincide with an eigenvalue $\omega^2_0$ of the differential operator
   \eqref{X''y}, and becomes arbitrarily small as $\omega^2 \to \omega^2_0$ (recall that 
   the latter means an instability of our static solution). Therefore, tracking the behavior of 
   the curve $X_{\rm num}(y; \omega)$ at the right end allows us to reveal the stability 
   and instability regions for various initial parameters of the system under consideration.

   Without loss of generality, we put $m=1$ and $X_0=1$, thus fixing the length scale for the
   problem and an insignificant factor in the solution for $X_{\rm num}$. Thus $r_b  $ remains 
   the only nontrivial changeable parameter of the~system.

\begin{figure}[H]    
\centering     
\includegraphics[width=7.5cm]{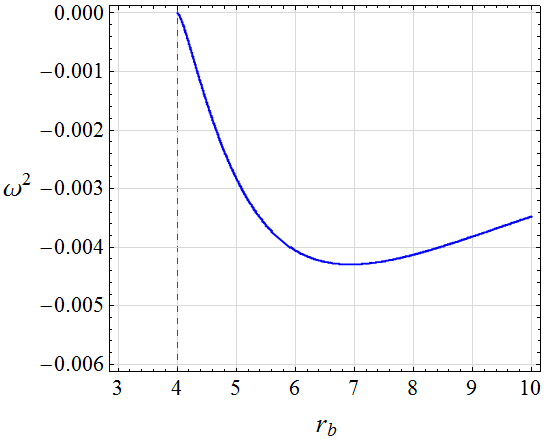}\qq
\caption{\small 
		The eigenvalue $\omega^2$ as a function of $r_b$. The red dashed line corresponds 
		to {$r_b = r_b^{\rm crit}\simeq 4.003996$}, which shows the right boundary of the 
		stability region.
		}     \label{omega-p-plot}
\end{figure}

  The results of our numerical analysis are presented in Figure \ref{omega-p-plot}. The plot 
  shows the existence of the eigenvalues $\omega^2$ as functions of $r_b  $ and reveals 
  the instability region at $r_b  > r_b^{\rm crit}$, where 
  {$r_b^{\rm crit}\simeq 4.003996\,m$} is a critical value of $r_b$ corresponding to the 
  limiting case $\omega^2=0$. At $r_b  < r_b^{\rm crit}$ the instability vanishes. The critical 
  value $r_b^{\rm crit}$ can be easily found numerically from the condition of vanishing of the 
  right-end value $X_{\rm num}(y_1)\bigr|_{\omega^2=0}\to 0$ as $r_b  \to r_b^{\rm crit}$.
  
\begin{figure}[H]   
\centering     
\includegraphics[width=7.5cm]{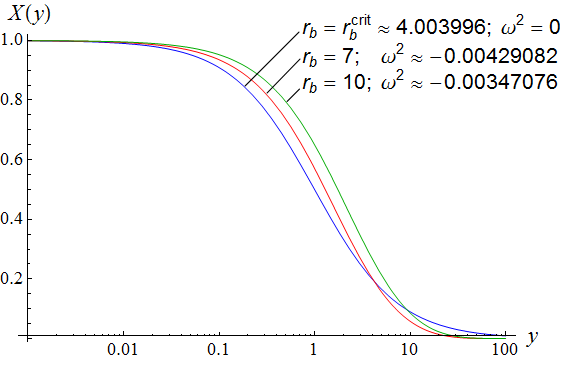}\qq
\includegraphics[width=7.5cm]{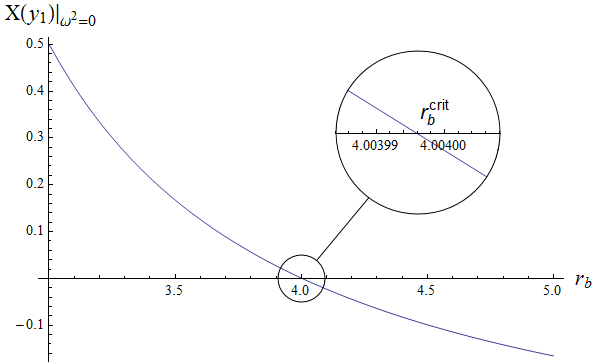}\qq
\caption{\small 
		(\textbf{Left}): numerical curves $X_{\rm num}(y)$ for various $\omega^2$ and $r_b  $. 
		(\textbf{Right}): vanishing of the right-end value $X_{\rm num}(y_1)\bigr|_{\omega^2=0}$ as 
		a function of $r_b  $ allows us to find the critical value {$r_b  ^{\rm crit}\simeq 4.003996$}.
		}     \label{Xnum-plots}
\end{figure}

  Figure \ref{Xnum-plots} (left panel) shows examples of numerical curves $X_{\rm num}(y)$ 
  for various $\omega^2$ and $r_b$, and also a method for seeking the critical value 
  $r_b^{\rm crit}$  (right panel). 
\vspace{-16pt}

   The stability condition $r_b < r_b^{\rm crit}$  can also be rewritten in terms of $m$ and $q$ 
   leading to a restriction on the ratio $q^2/m^2$. Combining it with the existence condition 
   $q^2 > 3m^2$,  we can finally write
\beq                              \label{range-ms}
		1 < \frac {q^2}{3 m^2} = \frac{r_b}{r_{\rm Scw}} \lesssim 2.001998.
\eeq 
  
  Within this range, in accord with \rf{k->q}, the parameter $k$ may be arbitrarily small, 
  in other words, the mirror surface radius may be very close to the would-be \Scw\ horizon.
  However, there is an observational constraint \cite{25-lim} on mirror surfaces according 
  to which very small differences $\ep = r_b  /r_{\rm Scw} -1 \lesssim 10^{-3}$ look unlikely. 

\subsection{Spherical Perturbations of Magnetic Black Holes}

  For the \bh\ solution \rf{mbh-r} we obtain the E-frame metric with 
\beq         \label{ds-bh}
		\e^{\xi(r)} = \frac{\sqrt{r-p}}{\sqrt{r}},
\qq
		\e^{2\ogamma} = \e^{-2\oalpha} = \frac{(r-2m)\sqrt{r-p}}{r^{3/2}},
\qq
		\e^{2\obeta} =  r^{3/2}\sqrt{r-p},
\eeq
   where $0< p < 2m$. Hence the function $U$ and the effective potential $\Veff$ are 
\bearr   			\label{Ubh}  
 		U(r) = \frac{2 p (8 m - 3 p)}{(4 r -3 p)^2 (r - p) (r -2 m)},
\yyy                         \label{Veff-bh}  
		\Veff(r) = \frac{r-2m}{16 (4 r -3 p)^2 (r - p) r^5}      
		\Big[378 m p^4 - 9 p^3 (184 m + 9 p) r + 128 p^2 (19 m + 3 p) r^2 
\nnn \inch	\inch		
				- 16 p (104 m + 27 p) r^3 + 128 (4 m + p) r^4		 \Big]. 
\ear

    The tortoise coordinate $z$ is connected with $r$ by the relations
\beq  		\label{rz-bh}
			z = \int \frac{r^{3/2} dr}{(r-2m)\sqrt{r-p}}, \qq
			z\Big|_{r\to 2m} \approx \frac{\log (r-2m)}{\sqrt{m-p/2}}, \qq
			r-2m\Big|_{z\to -\infty} \sim  \exp \big[z \sqrt{m-p/2}\big].
\eeq   
 
    Thus $z \approx r$ at large $r$, and $z\to -\infty$ as $r \to 2m$.
    The asymptotic behavior of $\Veff(r)$ is
\beq    
    			\Veff(r)\Big|_{r \to \infty}  = \frac{4m+p}{r^3} + O(r^{-4}), \qq
			\Veff(r)\Big|_{r \to -\infty} \approx \exp \big[z \sqrt{m-p/2}\big].
\eeq  
  
    The asymptotic behavior of solutions to \eqn{Schr} may be written as
\beq 		\label{Y-bh-as}
			Y(z) = C_5 \e^{i\omega z} + C_6  \e^{- i\omega z} \ \ (z\to \infty), \cm
			Y(z) = C_7 \e^{i\omega z} + C_8  \e^{- i\omega z} \ \ (z\to -\infty).
\eeq     

\begin{figure}[H]    
\centering     
\includegraphics[width=7.5cm]{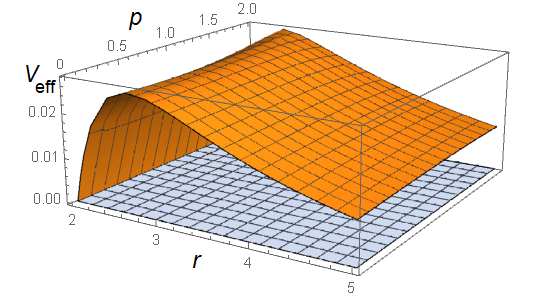}\qq
\includegraphics[width=7.5cm]{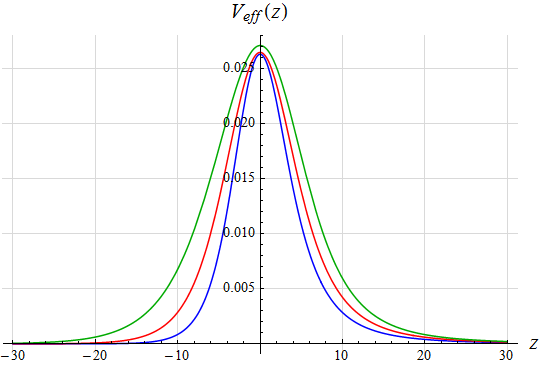}
\caption{\small 
		(\textbf{Left}): The effective potential for magnetic \bh\ \pbs\  \rf{Veff-bh} as a function of $r$ 
		and $p$.  The transparent level $\Veff =0$ shows that $\Veff > 0$ at all $r > 2m$. 
		{(\textbf{Right}): The effective potential as a function of $z$ for $p=0.1, 1, 1.5$ 
		(bottom--up).}
		}     \label{Vef-bh}
\end{figure}

{At both infinities $\Veff$ rapidly vanishes, and, as can be seen in Figure \ref{Vef-bh}, it is
    positive at all $r > 2m$ for any $p$. Addressing the stability problem for such \bhs\
    in the above manner, similarly to mirror stars, we come to a boundary-value problem 
    for \eqn{Schr} with zero boundary conditions and seek eigenvalues $\omega^2 \in \R$. 
    Then, the obtained $\Veff > 0$ implies the absence of negative or zero eigenvalues 
    $\omega^2$, consequently, these \bhs\ are stable under monopole \pbs\ at all values 
    of $p \in (0, 2m)$. 
    }

    \vspace{-6pt}
\vspace{-18pt}
    
{This result makes natural our next step: find out how the \bh\ perturbations decay if 
    allowed to lose their energy by radiating it either to infinity or to the horizon. This task 
    can be solved by methods widely used in numerous studies of quasinormal modes.
    Though, here we should stress that quasinormal modes characterize ``test'' fields in
    the background of a fixed particular metric, while we are here considering the system's 
    own \pbs\ of \mult\ origin, encompassing both scalar and metric disturbances.
    } 

\begin{figure}[H]    
\centering     
\includegraphics[width=14cm]{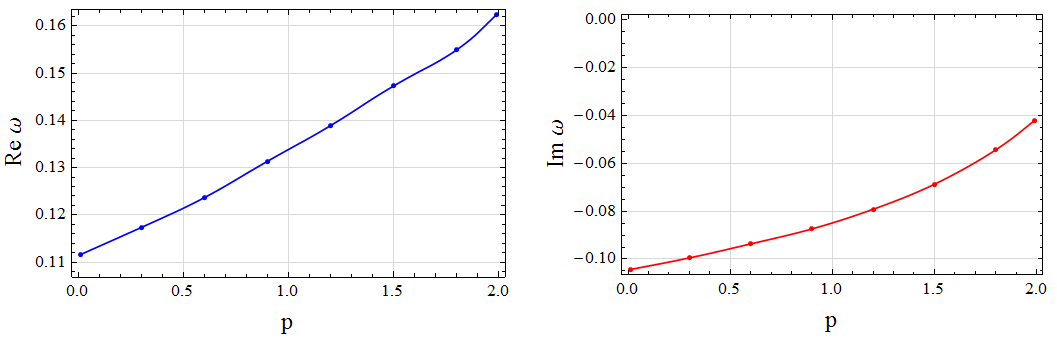}\qq
\caption{\small 
		{Graphic representations of the real and imaginary part of the WKB fundamental 
		mode $n=0$ interpolated between various values of $p$.}
		}     \label{WKBplot}
\end{figure}

\begin{figure}[H] 
\centering     
\includegraphics[width=7.5cm]{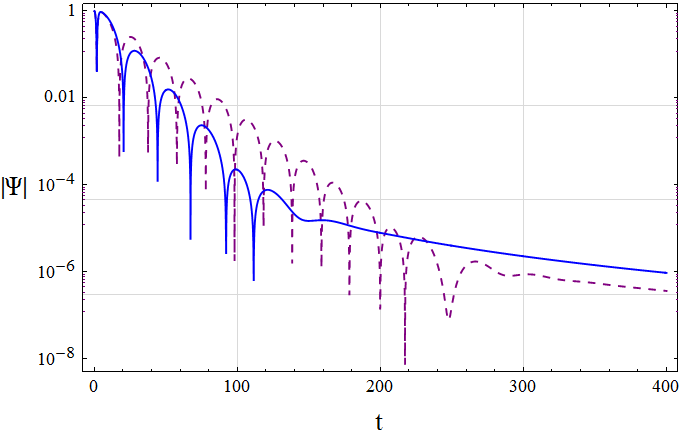}\qq
\caption{\small 
		{Example of the time domain profile for monopole perturbations with $p=1$ 
		(blue, solid) and $p=1.8$ (purple, dashed).}
		}     \label{TDplot}
\end{figure}

{
\subsection{Black Hole Perturbation Dynamics}

  To determine the monopole perturbation spectrum under realistic conditions,
  we consider the wavelike Equation~\eqref{wave} with typical boundary conditions, corresponding to purely ingoing waves at 
  the horizon and purely outgoing waves at spatial infinity. In terms of the tortoise coordinate, 
  these conditions are expressed as
\beq \label{QNMcond}
		 Y(z\to \pm \infty)\sim e^{\pm i\omega z},
\eeq
  where $\omega$ is a generically complex quantity whose real part describes the oscillation
  frequency while the imaginary part accounts for the damping rate.

  As can be seen from Figure~\ref{Vef-bh}, the effective potential $V_{\rm eff}(r)$ has a barrier 
  with a single peak and smoothly decays at infinity and at the event horizon without 
  any shape peculiarities. Such a good behavior of $\Veff$ allows us to apply well-known 
  numerical methods for calculation of the perturbations spectrum, widely used in the 
  literature~\cite{bh4}. Here we use two of them: the WKB approach and the Time Domain 
  Integration method. 

  The \textit{WKB (Wentzel--Kramers--Brillouin) approach} is based on expansion of the 
  solution to \eqn{Schr} in the so-called WKB series near the event horizon and at infinity in
  accordance with the boundary conditions, and matching these asymptotic series with the 
  Taylor expansion near the peak of the effective potential. This approach was originally 
  developed by Will and Schutz~\cite{Schutz} at the 1st WKB order and was extended to higher
  orders in the subsequent papers~\cite{Iyer, Kon2, Mat}. The general WKB formula for a 
  frequency $\omega^2$ can be written in the form~\cite{Kon3}
\beq
		\omega^2=V_0+A_2({\cal K}^2)+A_4({\cal K}^2)+...
		-i {\cal K}^2\sqrt{-2V_2}(1+A_3({\cal K}^2)+A_5({\cal K}^2)+...),
\eeq
  where each $k$-th order correction term $A_k$ is a cumbersome expression involving the 
  values of higher-order derivatives of $V_{\rm eff}(z)$ at its maximum; here by $V_0$ and 
  $V_2$ we denote the values of the effective potential and its second derivative at the 
  maximum.  There also exists an important improvement of the WKB method based on the 
  Pad\'e approximation technique~\cite{Mat,Hats}, which in most cases significantly increases 
  the accuracy.

  While the WKB approach is generically applicable for ``good'' shapes of the effective potential, 
  the convergence of the WKB series is asymptotic, i.e., it does not guarantee an increased 
  accuracy at each subsequent order. Moreover, this method works much better in the case 
  of higher multipole numbers $\ell\ge n$ (where $n$ is an overtone number) and is generally 
  not reliable in the case $\ell<n$. For $n=l$ the WKB method, especially improved by the 
  Pad\'e approximation, usually gives an acceptable accuracy that may be verified by comparison
  with other methods. We notice that the monopole perturbations  under consideration 
  correspond to $\ell=0$, so we restrict ourselves to analyzing the fundamental mode $n=\ell=0$ 
  for different values of the parameter $p$ (note that $p=0$ corresponds to the Schwarzschild 
  limit of the metric~\eqref{ds-bh}, 
  and $p=2$ to the extremal \bh\ \rf{ex-bh}, \rf{ex-bh-x}).

  In this particular study, we used the 9th order WKB method with $\tilde m=4$, where 
  $\tilde m$ is the polynomial degree in the corresponding Pad\'e approximant 
  expression~\cite{Mat, Kon3}. This choice is not strictly necessary, and it is chosen to illustrate 
  a good agreement with the results of another method based on time domain integration. 
  Our analysis is not aimed at ultra-precise calculation of the spectrum, but rather at showing 
  how typical approximate methods give a well-consistent result (within fractions of a percent), 
  revealing a clear qualitative behavior of the system in the damped oscillatory regimes. 
  For implementing the WKB procedure, we used the \textit{Mathematica}
  algorithm publicly 
  shared by the authors of~\cite{Kon3}.

  To obtain the time evolution profile of perturbations at some fixed $z$ (typically near the peak 
  of the effective potential), as well as to find the fundamental mode $n=0$, one can use 
  the so-called \textit{Time Domain integration}
  (TD) method which is in fact direct numerical
  integration of the wavelike Equation~\eqref{wave} performed in the null-cone variables 
  $u=t-z$, $v=t+z$ via the Gundlach-Price-Pullin discretization scheme~\cite{bh4, Gund}. 
  In this method, the interior of the future light cone is discretized by a rhombic grid with a 
  step $h$. Denoting the sides of a rhombic cell at the point $(u,v)$ as $S=(u,v)$, 
  $N=(u+h,v+h)$, $W=(u+h,v)$, $E=(u,v+h)$, the value of $\psi(N)$ can be calculated as
\beq
      \psi(N)=\psi(W)+\psi(E)-\psi(S)-h^2V_{\rm eff}(S)\frac{\psi(W)+\psi(E)}{8}+{\cal O}(h^4).
\eeq

  This scheme allows for obtaining a discrete ``time domain'' profile $\psi(t)$ at a fixed point 
  $z$ for a given initial perturbation, typically chosen as a Gaussian wave package on the 
  cone boundary~\cite{Gund}. The frequencies do not depend on the particular choice of this 
  initial perturbation, thus an obtained TD profile can be used to extract the fundamental mode. 
  For that, the Prony method is used~\cite{bh4} approximating the oscillatory part of a TD 
  profile by a sum of terms $\sim{C_j}e^{-i\omega_jt}$ within an appropriate ``fitting'' range 
  in $t$. In our case, the step $h=0.025$ was used, and the fitting range is about from 20 to 
  70, which yields a reliable oscillatory profile and the parameters of the extracted 
  fundamental mode.

  The results of our analysis are presented in Table\,\ref{tab1} and Figures \ref{WKBplot}, \ref{TDplot},
  where all quantities are expressed in units corresponding to our choice $m=1$.
  We see that the results obtained by the WKB-Pade and TD methods are in good agreement,
  mostly within the range of relative differences of a fraction of a percent (even though the case
  $n=\ell=0$ is not the best for applicability of the WKB method), and provide a clear picture 
  of the perturbations behavior with the changing parameter $p$. In particular, one can observe 
  that the real part (oscillation frequency) grows as $p$ increases, whereas the damping rate
  decreases, and the sensitivity of these quantities with respect to the change of $p$ is roughly
  comparable. One can also easily check that the limiting case $p\to 0$ restores the 
  Schwarzschild-type behavior of scalar perturbations. The other limiting case, the extremal 
  \bh\ solution \rf{ex-bh}, \rf{ex-bh-x} is also naturally included in the present setup.

  The conclusion made above on the stability of the \bh\ solution under consideration is directly
  verified by the time domain integration performed for various values of the parameter 
  $p\in(0,2m)$, so that the corresponding TD profile always represents damped oscillations
  indicating the system stability under monopole perturbations.
 }
 
 \vspace{-10pt}
\begin{table}[H]
\centering
\small
\caption{\small {The fundamental mode $n=0$ of \bh\ \pbs\ for various values of 
	$p$ calculated via the WKB Pad\'e and TD methods, with the corresponding relative 
	difference between them.}}	
\newcolumntype{C}{>{\centering\arraybackslash}X} 
\begin{tabularx}{\textwidth}{C>{\centering\arraybackslash}m{4.3cm}>{\centering\arraybackslash}m{4.3cm}cCC}
\toprule
 \boldmath{$p$} & \textbf{WKB9-Pad\'e} & \textbf{TD} & \boldmath{$\delta_{\text{Re,\%}}$} & \boldmath{$\delta_{\text{Im,\%}}$} \\
\midrule
0.01 & $0.111634-0.104343i$ 
& $0.111504-0.104535i$ & 0.12 & 0.18 \\
0.3  & $0.117382-0.099482i$ & $0.117460-0.099920i$ & 0.07 & 0.44 \\
0.6  & $0.123686-0.093657i$ & $0.124024-0.094124i$ & 0.27 & 0.50 \\
0.9  & $0.131363-0.087391i$ & $0.131315-0.087478i$ & 0.04 & 0.10 \\
1.2  & $0.138906-0.079178i$ & $0.139222-0.079318i$ & 0.23 & 0.18 \\
1.5  & $0.147281-0.068716i$ & $0.147572-0.068798i$ & 0.20 & 0.12 \\
1.8  & $0.154920-0.054365i$ & $0.155696-0.054439i$ & 0.50 & 0.14 \\
1.99 & $0.162377-0.042098i$ & $0.159158-0.042262i$ & 2.02 & 0.39 \\
\bottomrule
\end{tabularx}
\label{tab1}
\end{table}

\vspace{-20pt}

\vspace{-28pt}
    
\section{Concluding Remarks}\label{sec5}

{In this paper, we have determined the stability ranges of 5D magnetic mirror (topological)
  stars and \bhs\ under linear monopole \pbs. For mirror stars, this stability range is given by 
  \eqn{range-ms}, that is, roughly, $1 < q^2/(3m^2) < 2$,
  while \bhs\ turn out to be stable in the whole range $q^2/(3m^2) \leq 1$.
  At $q^2 > 6 m^2$, mirror stars experience an instability due to evolution of the extra 
  dimension whose metric coefficient $g_{55}$ behaves as a 4D scalar field. These results 
  evidently contradict some conclusions obtained by other authors. Thus, as asserted in 
  \cite{tops1}, ``topological stars are classically stable for the full range of parameters,'' 
  while ``magnetic black strings (i.e., \bhs) are free from classical linear instability for 
  $r_S/2 < r_B < r_S$,'' which in our notation means $1/2 < q^2/(3m^2) < 1$, contrary to 
  our stability conclusion for all $q^2 \leq 3m^2$.}
    
  On the other hand, in \cite{tops2, tops3}, the range of the ratio $r_B/r_S =q^2/(3m^2)$ 
  is divided into two sectors: Type I with $r_B/r_S > 3/2$, with a single photon sphere at 
  $r= r_B$, and Type II ($1 < r_B/r_S \leq 3/2$) containing two photon spheres $r = 3m$ 
  and $r = r_B$. In the introduction of \cite{tops3}, it is asserted that 
  ``the Type I sector was found to be free of instabilities,'' in contrast to our results: according 
  to \rf{range-ms}, the stability range covers the Type II sector but only partly the Type I sector.
  All these disagreements apparently require a further study.
  
{For the \bh\ solutions under consideration, we have also studied the monopole \pb\     
  dynamics, obtaining the spectrum of complex frequencies using the WKB and TD methods. 
  The results demonstrate an agreement between these two methods and also well confirm
  our stability conclusions. For the stable range of mirror star models, a similar study 
  cannot be carried out using the same standard methods which have to be modified owing to 
  occurrence of a regular mirror surface instead of a horizon, and we postpone such a study 
  for future work.}
  
  As follows from the regularity condition for mirror stars, their possible masses are related 
  to the compactification length $\ell$ of the extra dimension and are therefore constrained
  to be lighter than $\sim 10^{10}$\,g, see Section 2. Still one can notice that if we admit
  the multi-sheet nature of the extra dimension, then the regularity condition reads 
  $\ell = 2m/n$, and at possible large $n$ the mass $m$ may be much larger than $10^{10}$\,g. 
  Note the corresponding discussion of ``multiple bubbles'' in \cite{tops1}. 
  More than that, if we forget the regularity requirement and admit a conical-type 
  singularity (assuming that it must be somehow smoothed by quantum gravity effects), 
  the restriction on $m$ is no longer valid, while the mirror property for sufficiently light 
  particles, in whose wave functions the length scale is much larger than $\ell$, should be 
  preserved. In addition, in any case, close to $r = 2m$ or $r=r_b$ we are dealing with strong 
  gravitational fields and large curvatures, and it is necessary to consider QFT effects like those
  leading to Hawking \bh\ evaporation, which must be a subject of further studies. 
  
  Mirror stars of any size and mass and their possible clusters may be of interest from 
  an observational viewpoint, in particular, as possible dark matter candidates. 
  Even more interesting opportunities may emerge in the brane-world concept. For example,
  a particle may leave its brane with a certain probability and arrive at another brane.


%
%
%
%
%
%
%
%
%

	\small
	\bibliographystyle{scilight}
	
	

\end{document}